\documentclass[twocolumn,floatfix,prb,aps,showpacs]{revtex4}
\usepackage{graphicx,amsmath,amssymb,color}
\usepackage{nicefrac,multirow}
\usepackage[titletoc,title]{appendix}

\newcommand{\be}{\begin{equation}}
\newcommand{\ee}{\end{equation}}

\newcommand{\ba}{\begin{eqnarray}}
\newcommand{\ea}{\end{eqnarray}}

\begin{document}

\title{Spontaneous polarization of composite fermions in the $n=1$ Landau level of graphene}
\author{Ajit C. Balram,$^1$ Csaba T\H oke,$^2$ A. W\'ojs,$^3$ and J. K. Jain$^1$}

\affiliation{
   $^{1}$Department of Physics, 104 Davey Lab, Pennsylvania State University, University Park, Pennsylvania 16802, USA}
\affiliation{
   $^{2}$BME-MTA Exotic Quantum Phases ``Lend\"ulet" Research Group, Budapest University of Technology and Economics,
Institute of Physics, Budafoki \'ut 8, H-1111 Budapest, Hungary}
\affiliation{
   $^{3}$Department of Theoretical Physics, Wroclaw University of Technology, Wybrzeze Wyspianskiego 27, 50-370 Wroclaw, Poland}

\begin{abstract} 
Motivated by recent experiments that reveal expansive fractional quantum Hall states in the $n=1$ graphene Landau level and suggest a nontrivial role of the spin degree of freedom [Amet {\em et al.}, Nat. Commun. {\bf 6}, 5838 (2014)], we perform accurate quantitative study of the the competition between fractional quantum Hall states with different spin polarizations in the $n=1$ graphene Landau level. We find that the fractional quantum Hall effect is well described in terms of composite fermions, but the spin physics is qualitatively different from that in the $n=0$ Landau level. In particular, for the states at filling factors $\nu=s/(2s\pm 1)$, $s$ positive integer, a combination of exact diagonalization and the composite fermion theory shows that the ground state is fully spin polarized and supports a robust spin wave mode even in the limit of vanishing Zeeman coupling. Thus, even though composite fermions are formed, a mean field description that treats them as weakly interacting particles breaks down, and the exchange interaction between them is strong enough to cause a qualitative change in the behavior by inducing full spin polarization. We also verify that the fully spin polarized composite fermion Fermi sea has lower energy than the paired Pfaffian state at the relevant half fillings in the $n=1$ graphene Landau level, indicating an absence of composite-fermion pairing at half filling in the $n=1$ graphene Landau level. 
\end{abstract}
\maketitle

\section{Introduction}

The phenomenon of the fractional quantum Hall effect (FQHE)\cite{Tsui82} is the manifestation of one of the most remarkable many body states in nature. The electron's spin degree of freedom adds to the richness of this phenomenon, and experiments have demonstrated the existence of FQHE states with different spin polarizations as well as transitions between them as a function of the Zeeman energy \cite{Eisenstein89,Du95,Kukushkin99,Yeh99,Kukushkin00,Melinte00,Freytag01,Tiemann12,Liu14a}. Analogous transitions have been found in multi-valley systems, such as AlAs quantum wells \cite{Bishop07,Padmanabhan09} and two-dimensional electron system on an H-terminated Si(111) surface\cite{Kott14}, where, in the SU(2) limit, the valley index formally plays the same role as spin. 

The spin physics of FQHE is understood, qualitatively and semi-quantitatively, in terms of spinful composite fermions (CFs) \cite{Wu93,Park98,Park99,Park00b,Park01,Davenport12,Balram15}. A composite fermion \cite{Jain89,Lopez91,Halperin93,Jain07,Jain15} is the bound state of an electron and an even number ($2p$) of quantized vortices. Composite fermions form Landau-like levels called $\Lambda$ levels ($\Lambda$Ls) in a reduced magnetic field given by $B^*=B-2p\rho\phi_0$, where $\rho$ is the electron density and $\phi_0=hc/e$ is the flux quantum. The filling of composite fermions, $\nu^*$, is related to the electron filling by $\nu=\nu^{*}/(2p\nu^{*}\pm1)$. The prominently observed FQHE at the the fractions $\nu=s/(2ps\pm 1)$ is explained as the $\nu^*=s$ integer quantum Hall effect (IQHE) of composite fermions. For spinful composite fermions, we have $s=s_\uparrow+s_\downarrow$, where $s_\uparrow$ ($s_\downarrow$) is the number of occupied spin-up (spin-down) $\Lambda$Ls. This immediately predicts the allowed spin polarizations for any given fraction of this type, and the transition between two states with different spin polarizations occurs when CF $\Lambda$Ls of opposite spins cross as the Zeeman energy is varied. For the $\nu=s/(2s\pm 1)$ FQHE states, modeling them as weakly interacting composite fermions correctly predicts the energy ordering of the states, with the state with the smallest spin polarization having the lowest energy at small Zeeman energy and the fully spin polarized state winning at large Zeeman energies. 

There has been a resurgence of interest in FQHE in multicomponent systems as a result of the observation of well developed FQHE in multivalley systems of graphene \cite{Xu09,Bolotin09,Dean11,Feldman12,Feldman13,Amet15}. In particular, while the physics in the $n=0$ graphene Landau Level (GLL) is consistent with the expectation from weakly interacting composite fermions \cite{Feldman13}, experiments by Amet {\em et al.} \cite{Amet15} have indicated puzzling spin related behavior in the $n=1$ GLL. These authors have seen extensive FQHE in the $n=1$ GLL at the standard fractions of the form $\nu=s/(2s\pm 1)$. In addition, they have detected a change in the transport behavior in tilted field experiments up to magnetic fields of B$\sim$45 T. Such tilted field experiments are interpreted in terms of a variation of the Zeeman energy, and the experimental observations thus suggest that the spin degree of freedom remains relevant up to these magnetic fields.  This is surprising for the following reason. The relevant parameter for the spin physics is $\kappa=E_{\rm Z}/(e^2/\epsilon \ell)$, where $E_{\rm Z}$ is the Zeeman splitting, $\epsilon$ is the dielectric constant of the host material and $\ell=\sqrt{\hbar c/eB}$ is the magnetic length. In these experiments, the value of $\kappa$ at the highest magnetic fields is $\kappa\approx 0.07$ (assuming a $g$ factor of 2.0 and $\epsilon = 3.0$). In GaAs, where the spin physics of the FQHE has been investigated in detail\cite{Du95,Kukushkin99,Liu14}, the spin degree has been found to be relevant only for $\kappa \lesssim 0.02$\cite{Liu14}. It is unclear why the spin degree should remain relevant up to much higher values of $\kappa$ in graphene than in GaAs. That was our primary motivation for undertaking the study reported in this article. 

Previous theoretical work has addressed this issue, but not conclusively. Using the methods of density matrix renormalization group (DMRG), Shibata and Nomura \cite{Shibata09} have studied the FQH states in $n=1$ GLL in the torus and spherical geometries for systems up to 12 particles and surmised that the ground state at $\nu=2/3$ and 2/5 is fully polarized even at zero Zeeman energy. However, as we find below, the resolution to this issue is subtle and it requires much larger systems to conclusively determine the spin polarization of the ground state in the thermodynamic limit. T\H oke and Jain \cite{Toke07} used the methods of CF theory in the spherical geometry, and concluded that the ground state in the $n=1$ GLL is unpolarized at zero Zeeman energy and becomes fully polarized only at a finite Zeeman energy. This work employed an interaction in the $n=0$ LL that mimicked the Coulomb interaction in the $n=1$ GLL (because the relevant wave functions are most easily evaluated within the $n=0$ LL). However, in the course of our present study we have found that the interaction used in Ref.~\onlinecite{Toke07} was not sufficiently accurate for the purposes of spin physics, which is governed by very small energy differences. 

In this work we use a much more accurate interaction, and consider a wider range of filling factors than before. We also carry out a detailed comparison of the Moore-Read (MR) Pfaffian state and the CF Fermi sea (CFFS). We use methods similar to those in Refs. \onlinecite{Toke07} and \onlinecite{Toke12}.  Given that the observed fractions in the $n=1$ GLL belong to the standard $\nu=s/(2ps\pm 1)$ sequences, it appears very likely that the CF theory is qualitatively valid in the $n=1$ GLL, but it is a priori unclear how accurate it is quantitatively, and to what extent the residual interaction between composite fermions may be neglected.

Our conclusions, briefly, are as follows:

We find that the $\nu=s/(2s\pm 1)$ FQHE states in the $n=1$ GLL are fully polarized even at zero Zeeman energy. The situation is somewhat tricky for both 2/3 and 2/5, where the fully-spin-polarized and the spin-singlet states are almost degenerate insofar as their Coulomb interaction energy is concerned, and exact diagonalization studies are not able to decisively ascertain which of the two is the ground state. It is necessary to go to large systems using CF diagonalization (CFD) to determine which of the two has lower Coulomb interaction energy. Our calculations thus strongly suggest that the tilted field dependence observed in Ref.~\onlinecite{Amet15} is unrelated to the spin physics of composite fermions (and as yet not understood).

We further find that the fully spin polarized states are extremely accurately described in terms of composite fermions. The partially polarized states are also well described in terms of composite fermions. Taken together, these results imply that even though composite fermions are formed in the $n=1$ GLL, the model of weakly interacting composite fermions does not remain valid. The exchange interaction between composite fermions is strong enough to fully spin polarize the states. Analogous physics was found previously\cite{Park99} at fractions $\nu=s/(4s+1)$ in the $n=0$ LL. 

We also obtain the dispersion of the neutral excitons at $\nu=s/(2s\pm 1)$. For the spin polarized exciton, the dispersion is very similar to that in the $n=0$ LL\cite{Dev92,Scarola00}. The spin reversed exciton in the $n=1$ GLL does not show any ``sub-Zeeman energy" spin roton, as has been found for many FQHE states in the $n=0$ LL\cite{Murthy99,Mandal01,Wurstbauer11}. This is consistent with the above conclusion that the fully polarized state is stable against spin reversal. 

We have also considered filling factor 1/2 in the $n=1$ GLL and found that the CFFS remains stable to pairing. The MR Pfaffian state is not stabilized. This conclusion is consistent with previous work\cite{Toke07b,Wojs11a,Wojs11b}. Furthermore, we find that the Fermi sea also remains fully spin polarized even at zero Zeeman energy, in disagreement with the previous works\cite{Toke07b,Wojs11a,Wojs11b}. This behavior is to be contrasted with the $n=0$ LL where the CFFS with minimal spin polarization has the lowest interaction energy. 

Given that the some of the competing states are almost degenerate, one may suspect if LL mixing can play a crucial role in determining the ground state. As discussed previously in Ref.~\onlinecite{Balram15a}, a reliable treatment of LL mixing is a complicated task, especially because the LL mixing parameter is fairly large (greater than 1) in graphene. Nonetheless, we have estimated the corrections and found that the FQHE states remain fully spin polarized even when LL mixing is incorporated.

Throughout this work we shall neglect disorder, which may have less effect on these thermodynamic transitions than on single particle excitation gaps. Finite width effect are known to be significant for the spin physics of FQH states in GaAs\cite{Liu14}, but ought to be negligible in graphene. 

A discussion is in order regarding the validity of our use of the SU(2) model in the $n=1$ GLL. Much work has been done to understand the lifting of various degeneracies arising from the spin and valley degree of freedom in graphene\cite{Morpurgo06,Aleiner07,Dean11,Young12,Semenoff12,Roy14}. In principle, one can envisage the following three situations: 1) both the Zeeman and the valley-splitting terms are sufficiently small that the full SU(4) symmetry is preserved; 2) spin degeneracy is preserved while valley degeneracy is lifted or vice versa, whereby the appropriate symmetry would be SU(2); and 3) spin and valley degeneracies are both lifted. Experiments carried out at high magnetic fields (values similar to the ones of Amet {\em et al.}\cite{Amet15}) find that the $n=0$ GLL splits into four levels, thereby lifting both the spin and valley degeneracy, while in the $n=1$ GLL, only the spin degeneracy is lifted\cite{Zhang06}.

We assume in our work parameters such that the physics is well described within an SU(2) model, i.e., at most two of the four Landau bands are relevant. To the extent that the CF physics is valid, the neglect of SU(3) and SU(4) symmetries is clearly valid for states in which composite fermions fill one or two $\Lambda$ levels (e.g. 1/3, 2/5, 2/3), because the state with $s$ filled $\Lambda$ levels cannot exploit an SU($k$) symmetry for $k>s$. FQHE states at fillings factors with a numerator three or higher, such as 3/5, 3/7, 4/7, 4/9, {\em etc.}, can in principle support SU(3) and SU(4) spin structures\cite{Toke07}. However, we have found in the past\cite{Toke07} that if a ferromagnetic state is preferred in a two-component system, then the inclusion of further components does not change this feature. Hence, given that we will find a ferromagnetic state even for the SU(2) system in the $n=1$ GLL, our neglect of SU(3) and SU(4) symmetries is \textit{a posteriori} justified. (The only fraction in the $n=1$ GLL where a non-ferromagnetic Coulomb ground state is feasible is at the filling fraction 2/7, but even here, a tiny Zeeman energy fully polarizes the spin, leaving only the SU(2) valley degree to be considered here.)

The principal motivation of our work are the experimental results of Amet {\em et al.}\cite{Amet15}. It is clear that the SU(3) and SU(4) physics are not relevant for these experiments for the following two reasons: (i) The experimentally observed FQHE states in the $n=1$ GLL are precisely those expected from single component composite fermions, namely $s/(2ps\pm 1)$. (ii) The evolution of transport gaps in the presence of a parallel magnetic field suggests that the spin degree of freedom plays a role even up to very high values of the magnetic field. It is therefore sufficient to consider SU(2) physics for these experiments.

We use the language of ``spin" to label the two relevant indices, although our results below are also applicable if the two bands in question differ in their valley index (in that case ``spins" would refer to the valleys, and ``Zeeman splitting" to the splitting between valleys). At the end of the paper we briefly discuss the four component CFFS state which involves the full SU(4) symmetry of the combined spin and valley degree of freedom of graphene. Recently states involving the full SU(3) or SU(4) symmetry in graphene in the $n=0$ GLL have been discussed in Ref. \onlinecite{Wu15}. 

We note that the physics of the $n=0$ GLL, in the SU(2) limit, is identical to that of $n=0$ LL of non-relativistic electrons in GaAs, to the extent that finite width and LL mixing effects may be neglected. Our conclusions below for the $n=0$ LL thus apply to both the $n=0$ GLL and the lowest LL (LLL) of GaAs. 

The paper is organized as follows: In the next section we give a brief outline of the standard methods of exact and CF diagonlizations. We describe the effective interaction used to simulate the $n=1$ GLL in the $n=0$ GLL. Sections \ref{sec:gs_studies} and \ref{sec:ex_studies} are dedicated, respectively, to the energies of the ground state and low energy excitations at the filling factors $\nu=s/(2ps\pm 1)$. We include results from both CF theory and from exact diagonalization wherever these are available. In Section \ref{sec:Pf_vs_CFFS} we consider the question of stability of the CFFS state against the paired MR Pfaffian state. We end with conclusions in Section \ref{sec:conclusions}.

\section{Exact and CF diagonalization}
All our calculations are carried out in the spherical geometry \cite{Haldane83} wherein $N$ electrons reside on the surface of a sphere with a magnetic monopole of strength $2Qhc/e$ at its center. In this geometry the total orbital angular momentum $L$ is a good quantum number and ground states are seen to be uniform and incompressible i.e., have $L=0$ and have a finite gap to excitations. We consider states of all spin polarizations and apart from $L$ we also use the total spin angular momentum quantum number $S$ to characterize the states. The methods used in this work for exact and CF diagonalization are standard and details can be found in the literature\cite{Balram15,Jain97,Jain97b,Mandal02}. We briefly outline here some of the aspects that are relevant to the present work. 

\subsection{Exact diagonalization}
\label{sec:ed}
In the spherical geometry, we consider $N$ electrons confined to the surface of a sphere with the radial magnetic field produced by a Dirac magnetic monopole of strength $2Q\phi_{0}$. The $n$th LL corresponds to the angular momentum shell with $l=Q+n$ ($n=0$, $1$, $\dots$), which has $2l+1$ states labeled by the z-component of the angular momentum
$m=-l$, $-l+1$, \dots, $l$. The Coulomb interaction enters through the two-body matrix elements $\left<m_1,m_2|V|m_3,m_4\right>$ connected to the pseudopotential $V_\mu$ through the Clebsh-Gordan coefficients $C_{m_1m_2}^L\equiv\left<l,m_1;l,m_2|L,M=m_1+m_2\right>$:
\begin{equation}
\left<m_1,m_2|V|m_3,m_4\right>=\sum_{\mu} C_{m_1m_2}^{2l-\mu} V_\mu C_{m_3m_4}^{2l-\mu}
\end{equation}

In graphene the electron orbitals in all but the lowest LL are spinor states of the two-dimensional Dirac equation, with the two components corresponding to LL indices $n-1$ and $n$:
\begin{equation}
\left|\left|n,m\right>\right>=
\left(\left|n-1,m\right>\atop\left|n,m\right>\right).
\end{equation}
The difference in LL indices on the right hand side complicates complicates the derivation of the $(2l+1)$-fold LL degeneracy from the model of a spherical surface pierced by the magnetic flux. Specifically, when $Q$ defines a physical monopole, then the two components have different $l$ and hence different ranges of the allowed $m$'s. A common remedy has been to use pseudopotentials $V_\mu$ calculated (analytically) in the planar geometry \cite{Nomura06,Goerbig06,Apalkov06,Toke06,Toke07} truncated to the allowed range $\mu\le2l$. This will produce reliable results for sufficiently large systems, but may have strong finite size corrections.
We use an alternative approach following the direct analytical solution of the Dirac problem on a sphere by Jellal\cite{Jellal08}. Namely, we build the spinor wave function for graphene from a pair of orbitals with the same $l$, and thus with different $Q$. This leads to the expression for a spinor matrix element in graphene through the scalar matrix elements in GaAs, all derived in spherical geometry and hence all behaving properly at any range\cite{Toke06,Wojs11a}:
\begin{eqnarray}
&4&\left<\left<m_1,m_2||V||m_3,m_4\right>\right>\nonumber\\
&=&\left<n-0,m_1;n-0,m_2|V|n-0,m_3;n-0,m_4\right>\nonumber\\
&+&\left<n-1,m_1;n-0,m_2|V|n-1,m_3;n-0,m_4\right>\nonumber\\
&+&\left<n-0,m_1;n-1,m_2|V|n-0,m_3;n-1,m_4\right>\nonumber\\
&+&\left<n-1,m_1;n-1,m_2|V|n-1,m_3;n-1,m_4\right>.
\end{eqnarray}
We stress that all orbitals appearing in the above scalar matrix elements have the same $l$, which implies variable $Q=l-n$ or $l-n+1$, and that all scalar matrix elements are the spherical integrals. We first calculate the matrix element in units of $e^2/\epsilon R$, where $R$ is the radius of the sphere. The matrix element is then expressed in the units $e^2/\epsilon\ell$ where the magnetic length $\ell=R/\sqrt{Q}$ depends on the flux $2Q$. We choose the mean of the two flux values to define the magnetic length $\ell^{\text{av}}=R/\sqrt{Q^{\text{av}}}$, where $Q^{\text{av}}$ is defined as the average of the two flux values, i.e., $Q^{\text{av}}=[(l-n)+(l-n+1)]/2=l-n+1/2$ and express the energies in units of $e^2/\epsilon\ell^{\text{av}}$.
 
To take into account the presence of the positive background charge density we simply subtract $Ne^2/2\epsilon R$ from the per-particle electron-electron energy (adequate for systems with zero width\cite{Jain07}), where again we use $R=\ell^{\text{av}}\sqrt{Q^{\text{av}}}$. Finally, to incorporate the density correction we rescale the energy by multiplying the background subtracted per-particle energy by the factor of $\sqrt{2\nu Q^{\text{av}}/N}$. We note that this scheme produces correct pseudopotentials for large values of the relative angular momentum $\mu$, an approximately linear dependence of the per-particle ground state energies on $1/N$, and identical thermodynamic energies as obtained using truncated planar pseudopotentials.

\subsection{Effective interaction}
\label{sec:eff_int}
\begin{figure}
\begin{center}
\includegraphics[width=0.5\textwidth]{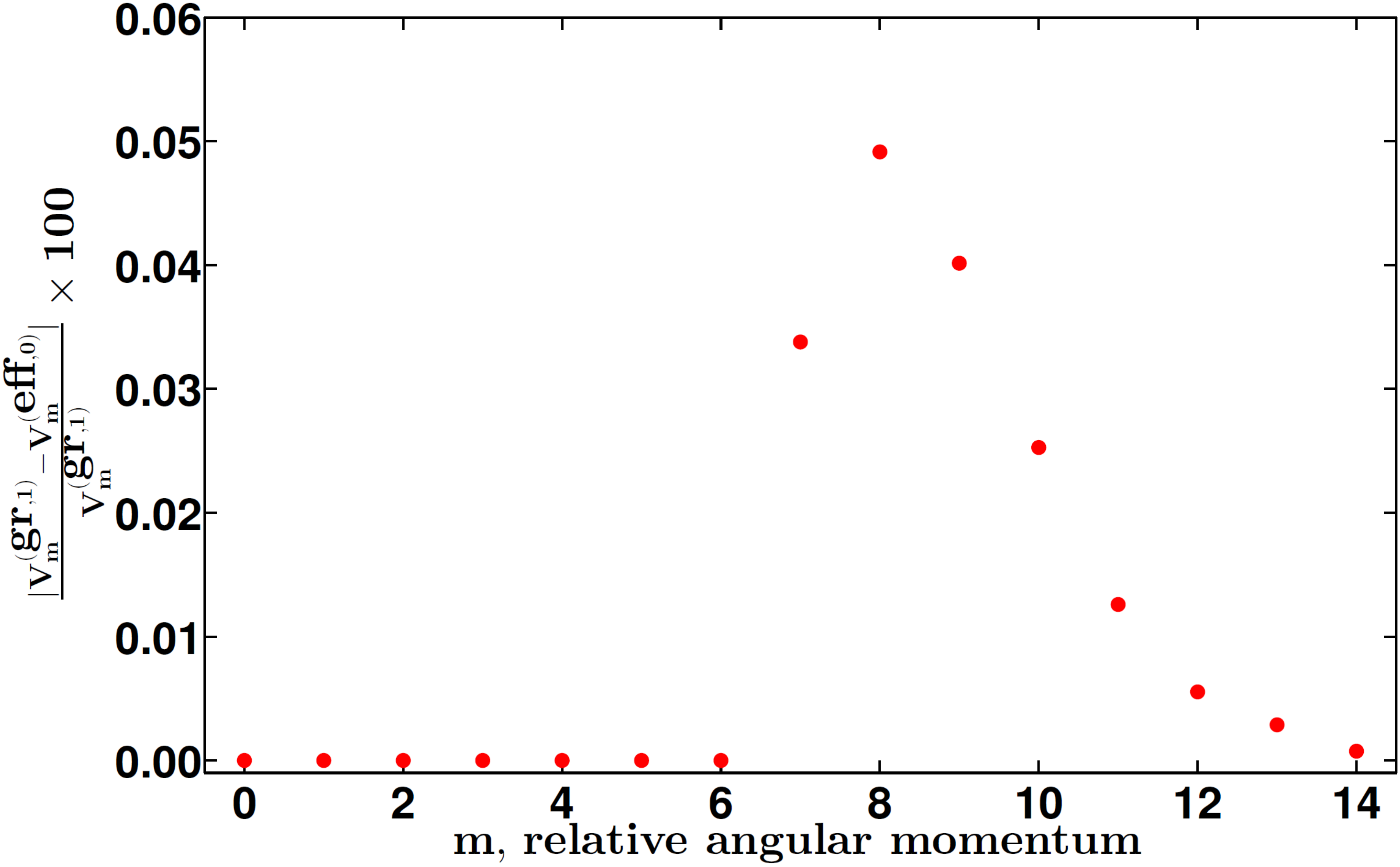}
\caption{(color online) Comparison between the exact pseudopotentials for the Coulomb interaction in the $n=1$ GLL and the pseudopotentials of the effective interaction given in Eq. \ref{eq:eff_int} in the $n=0$ GLL in the disc geometry.}
\label{fig:pps}
\end{center}
\end{figure}

The problem of interacting electrons in higher LLs is mathematically equivalent to a problem of electrons restricted to the LLL of GaAs (or the $n=0$ GLL) but interacting via an effective interaction $V^{\text{eff}}(r)$ which reproduces the higher LL Haldane pseudopotentials \cite{Haldane83}. This turns out to be the most efficient way of doing calculations within the CF theory, as the CF wave functions are most easily evaluated in the $n=0$ GLL. In the spherical geometry the pseudopotentials depend on the size of the sphere. Consequently, one must find an effective interaction for each system size. To circumvent this problem we take the pseudopotentials from the disc geometry and use them for the spherical geometry. This procedure should give the correct results in the thermodynamic limit, and we carefully evaluate below the thermodynamic limits for all our CF calculations. The exact Coulomb pseudopotentials of graphene on the disc geometry are well known (see for example \cite{Nomura06,Goerbig06,Apalkov06,Toke06,Toke07}). For the $n=1$ LL these are given by:
\begin{equation}
V^{(\text{gr},1)}_{m}=\Big[\frac{1}{r}\Big]^{(\text{gr},1)}_{m}= \Big( m^2-\frac{15m}{8} + \frac{153}{256}\Big) \frac{\Gamma(m-\frac{3}{2})}{2\Gamma(m+1)} 
\label{eq:V1_coulomb}
\end{equation}
The following form for the effective interaction in the lowest LL is used to simulate the physics of $n=1$ GLL \cite{Shi08,Toke06a}:
\begin{equation}
V^{\text{eff}}(r)=\frac{B_1}{r}+\frac{B_3}{\sqrt{r^6+1}}+\frac{B_5}{\sqrt{r^{10}+10}}+\sum_{k=0}^{k=6}C_{2k}r^{2k} \exp(-r^2)
\label{eq:eff_int}
\end{equation}
where $r$ is the distance in units of the magnetic length $\ell$. The coefficients $B_{i}$'s and $C_{i}$'s are listed in Table \ref{tab:coeff_eff_int}. $B_{1}$, $B_{3}$ and $B_{5}$ are fixed by the long range part of the Coulomb interaction while the $C_{i}$'s are obtained by matching the first seven pseudopotentials of $V^{(\text{gr},1)}_{m}$ with those of the effective interaction. We find that fixing more than the first seven pseudopotentials leads to a highly oscillating form of the real space interaction which is undesirable, while fixing fewer pseudopotentials increases the error in the remaining pseudopotentials. Fig.~\ref{fig:pps} shows the comparison of the pseudopotentials of this effective interaction in the lowest LL with those of the Coulomb pseudopotentials in the $n=1$ GLL in the disc geometry. We see that the difference between the two is less than 0.06\% for all values of $m$. We use this effective interaction to study large systems in the spherical geometry. This interaction is more accurate than the effective interaction used in a previous study by some of the authors\cite{Toke07,Toke07b}, and the results below supersede the previous results for the $n=1$ GLL.\\

\begin{table}
\begin{center}
\begin{tabular}{|c|c|}
\hline
\multicolumn{1}{|c|}{Coefficient} & \multicolumn{1}{|c|}{Value} \\ \hline
$B_1$ 	 			  & $1.0$  		\\ \hline
$B_3$ 	 			  & $0.5$  		\\ \hline
$B_5$ 	 			  & $0.5625$		\\ \hline
$C_0$ 	 			  & $42.55210599$	\\ \hline
$C_2$ 	 			  & $-284.7958520$ 	\\ \hline
$C_4$ 	 			  & $376.9514252$ 	\\ \hline
$C_6$ 	 			  & $-179.3196140$ 	\\ \hline
$C_8$ 	 			  & $36.58286249$ 	\\ \hline
$C_{10}$ 			  & $-3.232277646$ 	\\ \hline
$C_{12}$ 			  & $0.1004240132$       \\ \hline
\end{tabular}
\end{center}
\caption{Coefficients of Eq. \ref{eq:eff_int} which produces the effective interaction used to simulate the physics of the $n=1$ GLL in the lowest LL.} 
\label{tab:coeff_eff_int} 
\end{table}

\subsection{CF theory}
\label{sec:CF_theory}
The CF theory states that the system of strongly interacting electrons in a magnetic field at a filling factor $\nu=s/(2ps\pm1)$ can be mapped into a system of weakly interacting composite fermions at a filling factor $\nu^{*}=s=s_{\uparrow}+s_{\downarrow}$, where $s_{\uparrow}$ ($s_{\downarrow}$) is the filling of spin-up (spin-down) composite fermions\cite{Jain89}. We denote this state by $(s_{\uparrow},s_{\uparrow})$. The ground state at these special filling factors is described by the CF wave function:
\be
\Psi_{\frac{s}{2ps\pm 1}}={\cal P}_{\rm LLL} \Phi_{\pm s} J^{2p}
\label{Jain_wf}
\ee
where 
$
J= \prod_{1\leq j<k \leq N}(z_j-z_k)
$
is the so-called Jastrow factor, $z_{i}$ denotes the coordinate of the $i^{th}$ electron, $\Phi_{s}=\Phi_{s_{\uparrow}}\Phi_{s_{\downarrow}}$ is the Slater determinant of $s$ ($s_{\uparrow}$ spin-up and $s_{\downarrow}$ spin-down) filled LLs for electrons and $\Phi_{-s}=[\Phi_{s}]^*$. The states at $\nu=s/(2ps-1)$ require reverse flux attachment, and describe composite fermions in an effective field that points in a direction antiparallel to the applied magnetic field $B$. These wave functions involve LLL projection, for which three slightly different schemes have been employed. The most accurate one is the so called ``hard core" projection, which has been shown to be very accurate \cite{Wu93}, but is not easy to implement. We will therefore not consider the hard-core projection here. Even for the non-hard-core projection, two methods have been employed, namely direct projection\cite{Dev92,Wu93} and the Jain-Kamilla projection \cite{Jain97,Jain97b,Jain07,Moller05,Davenport12}. Here, for technical reasons, we will exclusively consider the wave functions with Jain-Kamilla projection, which produces less accurate results. The details of the projection method have been outlined in the literature \cite{Jain97,Jain97b,Jain07,Moller05,Davenport12} and will not be repeated here.\\

In addition to the above wave functions for the incompressible states at $\nu=s/(2ps\pm 1)$, we also study trial wave functions for the state at $\nu=1/2$. For the CFFS, we will consider wave functions of the form in Eq.~\ref{Jain_wf} for which composite fermions experience zero flux \cite{Rezayi94} and then take the thermodynamic limit to obtain the energy of the state. We also study the  Pfaffian wave function\cite{Moore91} ${\rm Pf}[M_{ij}] J^2$, where Pf denotes the Pfaffian and $M_{ij}=(z_i-z_j)^{-1}$, which describes a chiral p-wave paired state of composite fermions,\cite{Read00} and is the most promising candidate for the FQHE state at $\nu=5/2$ in GaAs systems.

For the CFFS, we have considered for $\Phi(B^*=0)$ a single component state (which approaches in the thermodynamic limit a fully spin polarized CFFS), a two component state with equal number of particles in both components (spin singlet CFFS) and a four component state with equal number of particles in all four components (SU(4) singlet CFFS).  In all cases, we work with filled shell states, so the state is uniform and isotropic. 

The total energies also include contributions from the positive neutralizing background, which we take into account under the assumption that a positive charge of $Ne$ is uniformly distributed on the surface of the sphere. The net energy of $N$ particles is given by: $E_{N}=E_{el-el}-\frac{N^2}{2\sqrt{Q}}\frac{e^2}{\epsilon\ell}$, where the first term gives the contribution from electron-electron interaction, which we evaluate using the methods of CF theory, and the second term takes into account the electron-background and background-background interactions. To make a comparison with experiments we need to find the energies of the ground states in the thermodynamic limit. The density for a finite system depends on the number of particles $N$ and is different from its thermodynamic limit. Therefore, to reduce the $N$ dependence we first apply a density correction to the energy, redefining it to be \cite{Morf86}: $E_{N}^{'}=(\frac{2Q\nu}{N})^{1/2}E_{N}$ and then extrapolate to $N\rightarrow \infty$. We note that the energies extrapolated to the thermodynamic limit with and without the density corrections are slightly different from each other. All numbers quoted in this work include density correction. 

Note that the background subtraction for exact diagonalization is done slightly differently from the procedure mentioned here. For exact diagonalization the systems considered are not very large, so a more careful consideration is required while doing the background subtraction. For results obtained using CF theory we go to much larger systems whereby the difference from these $\mathcal{O}(1)$ corrections are negligible. Also since we take the thermodynamic limit, the results are expected to be independent of these $\mathcal{O}(1)$ corrections. We also note that the energies for a finite system obtained below from exact diagonalization and CF theory are not directly comparable since the pseudopotentials of the effective interaction for a finite sphere are different from the $n=1$ GLL Coulomb disk and spherical pseudopotentials used in the exact diagonalization. The two results, however, should match in the thermodynamic limit.

We evaluate energies of the wave functions given in Eq. \ref{Jain_wf} using the Metropolis Monte Carlo method where in we sample with the above wave functions and use the $V^{\text{eff}}(r)$ interaction as the Hamiltonian. These energies are denoted by ``CF w.f''. and are given in Tables \ref{tab:grn_2} and \ref{tab:grn_3}. To improve on the above wave functions we use the method of CF diagonalization which is described next. 

\subsection{CF diagonalization \cite{Mandal02}}
\label{sec:cfd}
We take the $V^{\text{eff}}(r)$ interaction as the Hamiltonian and evaluate the energies of the states using the Metropolis Monte-Carlo method as follows: First, construct simultaneous eigenstates of the $L^{2}$ and $S^{2}$ operators in the corresponding IQHE system. These states are then multiplied by the factor of $J^{2p}$ and projected onto the lowest LL, a procedure known as composite-fermionization. This gives us the required $(L^2,S^2)$ eigenstates since $J^2$ commutes with both these operators. The set of basis states $\{\Psi_{i}\}$ are obtained by taking all possible $(L^2,S^2)$ eigenstates. Evaluation of the Hamiltonian matrix in this basis involves multi-dimensional integrals for which we use the Metropolis Monte-Carlo method. In general, CF wave functions are not orthogonal to each other, therefore the Gram-Schmidt orthogonalization procedure is implemented to find the final eigenenergies. To improve the efficiency of the computation we do the calculation within the subspace of $(L^2,S^2)$ eigenstates. As was stated above the lowest LL projection is carried out using the Jain-Kamilla method details of which can be found in the literature \cite{Jain97,Jain97b,Jain07}.

\begin{figure}
\begin{center}
\includegraphics[width=0.5\textwidth]{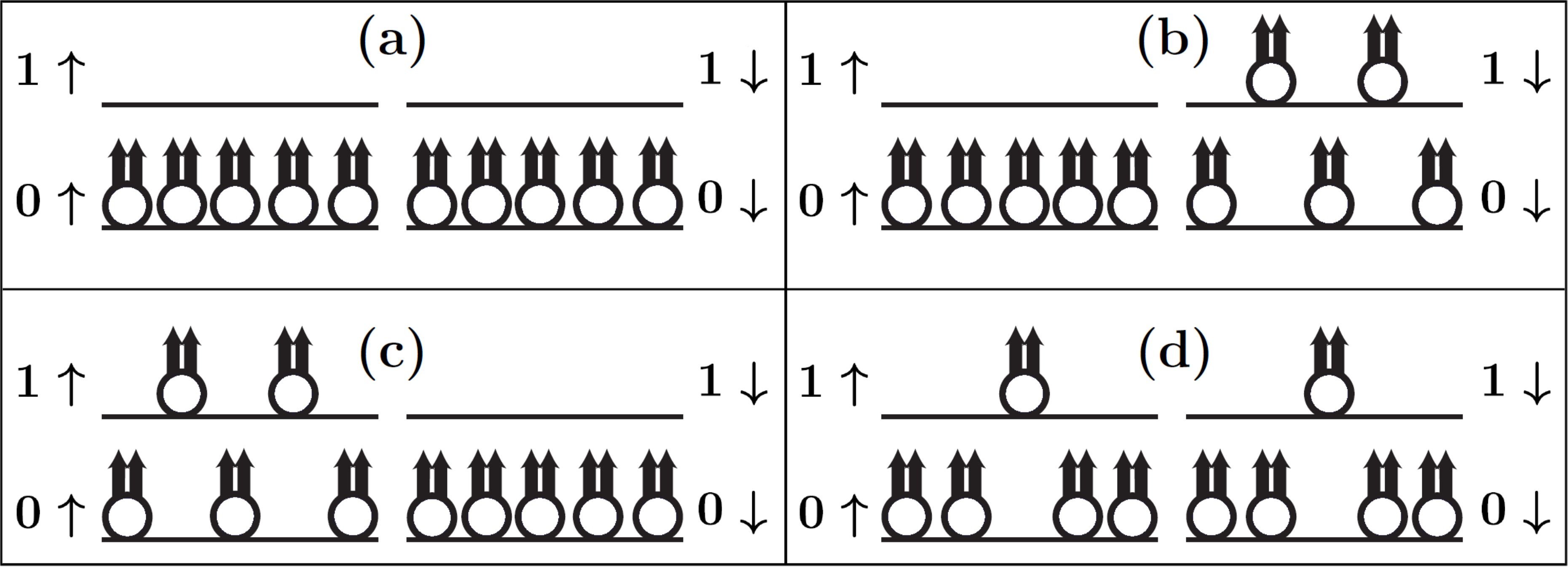}
\caption{2/5 spin-singlet states. Panel (a) shows the schematic of the the 2/5 spin-singlet CF state. At the first order approximation, composite fermion diagonalization mixes the state in (a) with the states shown in panels (b), (c) and (d) to obtain a new ground state with lower energy than the state in (a). Reproduced from Ref.~\onlinecite{Balram15a}.}
\label{fig:1}
\end{center}
\end{figure}

The ground state wave functions can be improved by including CF excitons (where a single CF exciton is a pair of CF particle and hole) in the basis and performing CFD \cite{Mandal02}. A single CF exciton does not couple to the ground state because its smallest angular momentum is one. Therefore, we need a minimum of two CF excitons to improve the ground state. In Fig. \ref{fig:1} we show the excitations we considered to improve the 2/5 spin-singlet state. The Hilbert space grows very quickly with the number of excitons included in the basis for CFD, so we restrict ourselves to at most two excitons. Among the ground state wave functions shown above, the fully polarized ones are extremely accurate, so this procedure of including two CF excitons in the basis of CFD only marginally improves the ground state energy of the fully polarized state. For the partially polarized states the improvement is more significant -- for the spin-singlet states the energies improve by around $10\%$. (Note that the CF excitons in which the constituent CF particle and CF hole are separated by two $\Lambda$Ls have $L\geq 2$ and therefore do not renormalize the incompressible ground state. Spin-flip excitons produce states in the degenerate $(2S+1)$ multiplet and therefore have identical ground state energy but different $S_{z}$ value. Consequently the two aformentioned excitons are not be included in the Hilbert space for CFD.) We have performed CFD for states at 2/5 and 3/7. 

\section{Ground state spin polarization}
\label{sec:gs_studies} 

We have considered the spin physics at many fractions. For the sequences $\nu=s/(2s \pm 1)$, we have investigated 2/5, 3/7, 4/9, 2/3, 3/5, 4/7; for the sequences $\nu=s/(4s \pm 1)$, we have studied 2/9, 3/13, 2/7, 3/11; and for the sequences $\nu=s/(6s \pm 1)$ we have studied 2/13, 2/11, 3/17. As discussed below in more detail, all states at $\nu=s/(2s\pm 1)$ are fully spin polarized, although some non-fully spin polarized states are possible for the states of composite fermions carrying four or six vortices.

\subsection{Exact diagonalization}
\begin{figure}
\begin{center}
\includegraphics[width=0.5\textwidth]{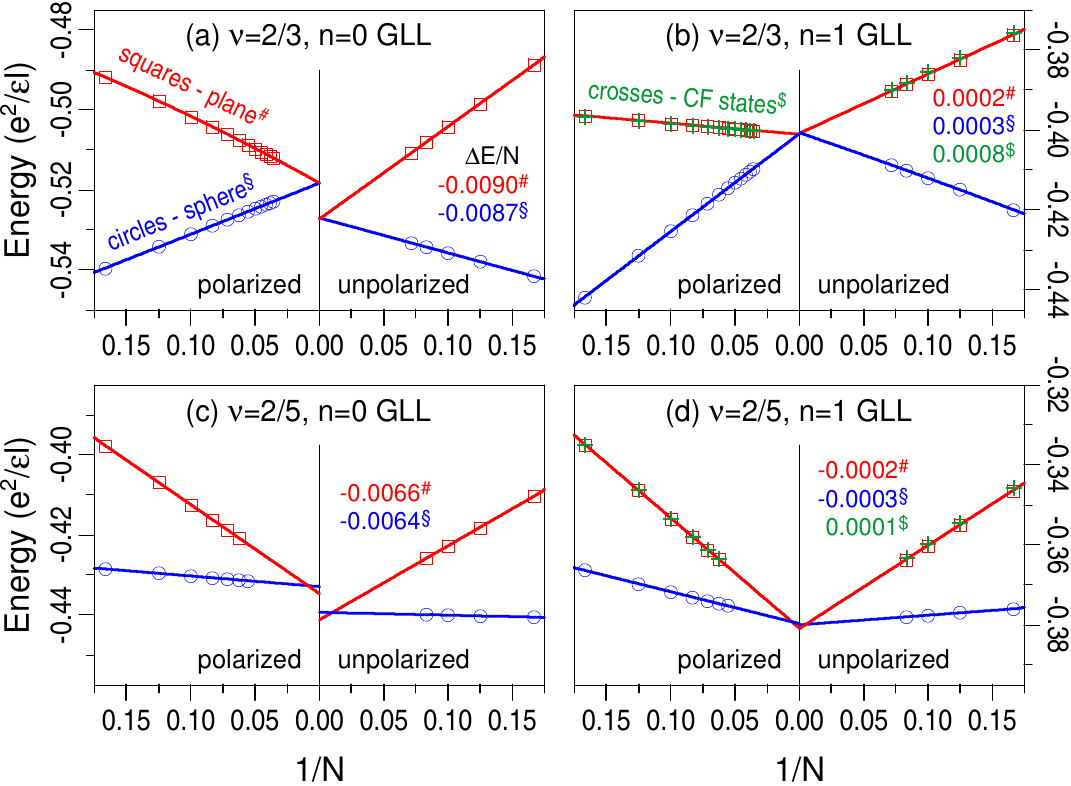}
\caption{(color online) Thermodynamic extrapolation of the ground state energies, obtained by exact diagonalization, for the fully polarized and spin-singlet states in the $n=1$ GLL at filling factors 2/3 (panel b) and 2/5 (panel d). Also shown for comparison are the corresponding results in the $n=0$ GLL in panels a (2/3) and c (2/5). We show results obtained from disk (red squares) and spherical pseudopotentials (blue circles). The background subtraction and density correction is carried out as described in Section \ref{sec:ed}. Additionally, with the green crosses we show the expectation values of the $n=1$ GLL Coulomb interaction with respect to the exact $n=0$ GLL ground states; the latter are known to be nearly exactly described by the wave functions of CF theory given in Eq. \ref{Jain_wf}. All energies are quoted in units of $e^2/\epsilon\ell$.}
\label{fig:exact_gs_n_2}
\end{center}
\end{figure}

\begin{figure}
\begin{center}
\includegraphics[width=0.5\textwidth]{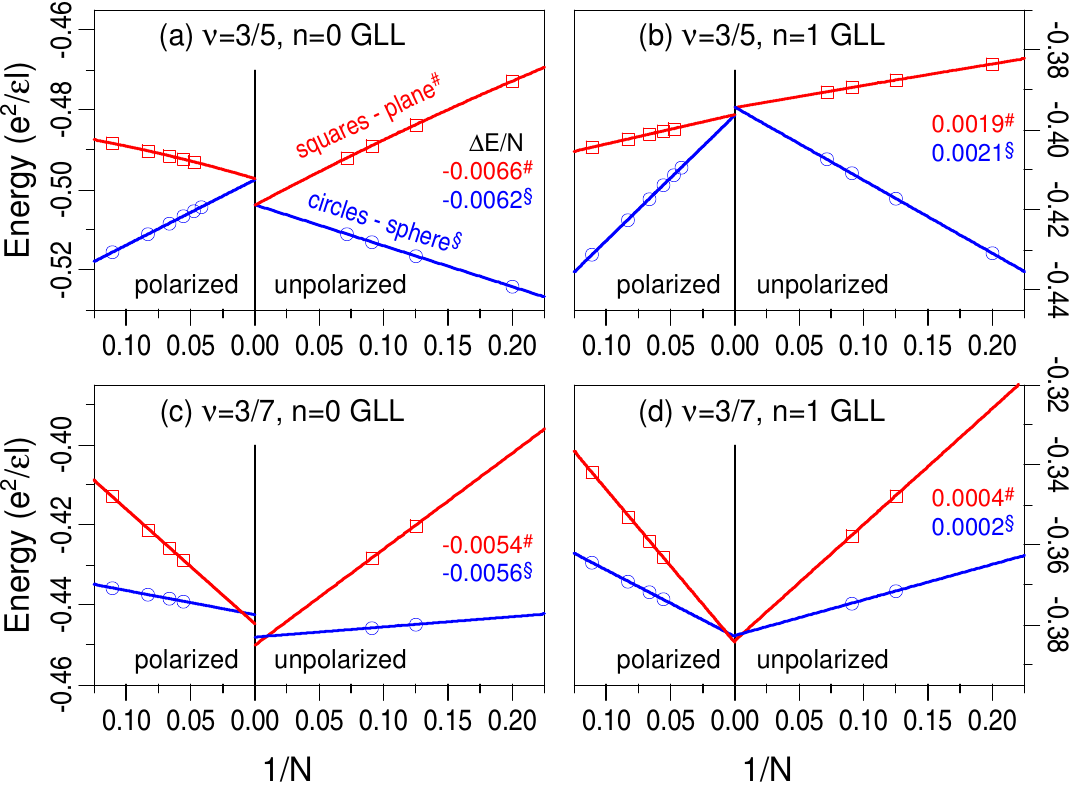}
\caption{(color online) Same as in Fig. \ref{fig:exact_gs_n_2} but for $\nu=3/5$ and $\nu=3/7$.}
\label{fig:exact_gs_n_3}
\end{center}
\end{figure}

We have performed exact diagonalization for up to $N=28$ ($N=14$) for fully polarized (spin-singlet) 2/3 and $N=16$ ($N=12$) for fully polarized (spin-singlet) 2/5. Similarly we have also carried out exact diagonalization up to $N=24$ ($N=14$) for fully polarized (partially polarized) 3/5 and $N=18$ ($N=11$) for fully polarized (partially polarized) 3/7. The results are shown in Fig.~\ref{fig:exact_gs_n_2} and \ref{fig:exact_gs_n_3} where we use both the truncated disk and the spherical pseudopotentials. Both pseudopotentials converge to the same thermodynamic limit, explicitly confirming the validity of our procedure.  We show here only the contribution of the Coulomb interaction energy; the Zeeman contribution can be incorporated separately. For the $n=0$ GLL, the spin singlet state has lower Coulomb energy, consistent with previous calculations \cite{Wu93,Park98}. However, in the $n=1$ GLL, the fully spin polarized state and the spin singlet state are almost degenerate in their Coulomb interaction energy for both 2/3 and 2/5. This is a surprising result, but may appear somewhat more natural within the CF theory, where both of these states map into filling factor $\nu^*=2$ of composite fermions. Because there is no exact symmetry relating the energies of the spin singlet and fully spin polarized states, the near degeneracy in their interaction energy is accidental, and larger systems should select one of the two as the thermodynamic ground state. Our calculations below based on the CF theory and also our results on the spin wave dispersion strongly point to the fully spin polarized state having slightly lower energy than the spin singlet state at both 2/3 and 2/5. The analysis of 3/5 and 3/7 (which both map to $\nu^*=3$ of composite fermions) is similar. Exact diagonalization suggests that the fully polarized and partially polarized states at these filling factors are almost degenerate, and the study of larger systems using the CF theory favors the fully polarized state.

In Table \ref{overlaps}, we show the overlaps between the ground states of the $n=0$ and $n=1$ GLL for different spin polarized states at various filling factors. The fully spin polarized states in the two Landau levels are nearly identical. For the partially spin polarized or the spin singlet states, the overlaps between the $n=0$ and $n=1$ GLL states are not extremely high, but still high. We have also calculated the expectation value of the $n=1$ GLL Coulomb interaction with respect to the exact ground state of the $n=0$ GLL.  These are shown as green crosses in Fig. \ref{fig:exact_gs_n_2} and agree almost perfectly with the red squares. Because the $n=0$ GLL states are nearly exactly captured by the CF theory, these comparisons give us confidence in the applicability of the CF theory for the states in the $n=1$ GLL as well. Note that this is very different from the behavior in GaAs, where the states in the $n=1$ LL are substantially different from those in the $n=0$ LL.  

\begin{table*}
\centering
\begin{tabular}{|c|c|c|c|c|c|c|c|c|c|c|}
\hline 
$(s_\uparrow,s_\downarrow)$ & $\nu$ & $N_{\uparrow}$ & $N_{\downarrow}$ & $2Q$ & dimension & $|\langle \psi_{S}^{(1)}|\psi_{S}^{(0)} \rangle|$  & $|\langle \psi_{D}^{(1)}|\psi_{D}^{(0)} \rangle |$ & $|\langle \psi_{S}^{(0)}|\psi_{D}^{(0)} \rangle |$ & $| \langle \psi_{S}^{(1)}|\psi_{D}^{(1)} \rangle |$ \\ \hline
(2,0) & 2/5 & 16 & 0 & 36 &  155484150 &0.9998984626    &0.9999167952    		&0.9998458746 &0.9998960866\\
(1,1) & 2/5 &  6 & 6 & 27 & 2211680688 &0.9313810788    &0.9516235092$^{\dagger}$	&1$^{\dagger}$&0.9965515722\\
(2,0) & 2/3 & 26 & 0 & 39 &  259140928 &0.9998146190    &0.9998485003    		&0.9998260141 &0.9998807588\\
(1,1) & 2/3 &  7 & 7 & 20 &  280934870 &0.7882484953    &0.9522359035    		&0.9999709236 &0.9279167113\\
(3,0) & 3/7 & 18 & 0 & 37 &  386905330 &0.9999276937    &0.9999272643    		&0.9997009362 &0.9997601209\\
(2,1) & 3/7 &  8 & 3 & 22 &   17969272 &0.9133546180    &0.9374707945    		&0.9998133677 &0.9981942320\\
(3,0) & 3/5 & 21 & 0 & 36 &  155484150 &0.9998984626    &0.9999167952  	  		&0.9998458746 &0.9998960866\\
(2,1) & 3/5 & 10 & 4 & 23 &  383215178 &0.8480598539    &0.9209952153   	 	&0.9998273715 &0.9819871145\\ \hline
\end{tabular} 
\caption{\label{overlaps} Table shows the Hilbert dimension and overlaps of the ground states of the $n=0$ [superscript (0)] and $n=1$ [superscript (1)] GLL calculated using the spherical [subscript S] and truncated disc [subscript D] pseudopotentials for the largest systems considered in this work. $^{\dagger}$For the 2/5 spin singlet state listed above, we do not have the vector $|\psi_{D}^{(0)} \rangle$. We presume it is almost identical to $|\psi_{S}^{(0)} \rangle$ and quote the numbers $|\langle \psi_{D}^{(1)}|\psi_{S}^{(0)} \rangle |$ and $|\langle \psi_{S}^{(0)}|\psi_{S}^{(0)} \rangle |$.}
\end{table*}

\subsection{CF wave functions}
\begin{figure*}[htbp]
\begin{center}
\includegraphics[width=0.64\columnwidth,keepaspectratio]{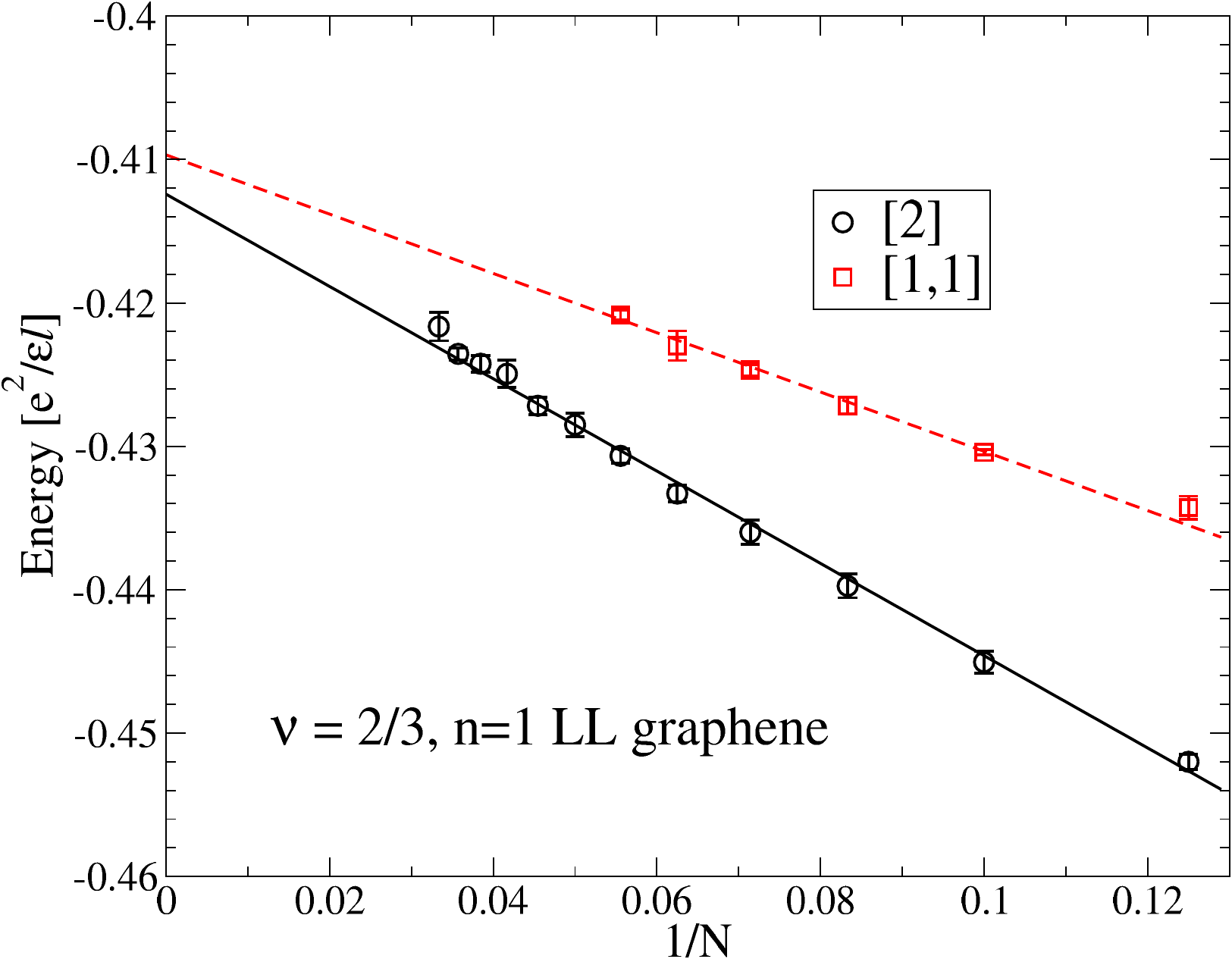}
\includegraphics[width=0.64\columnwidth,keepaspectratio]{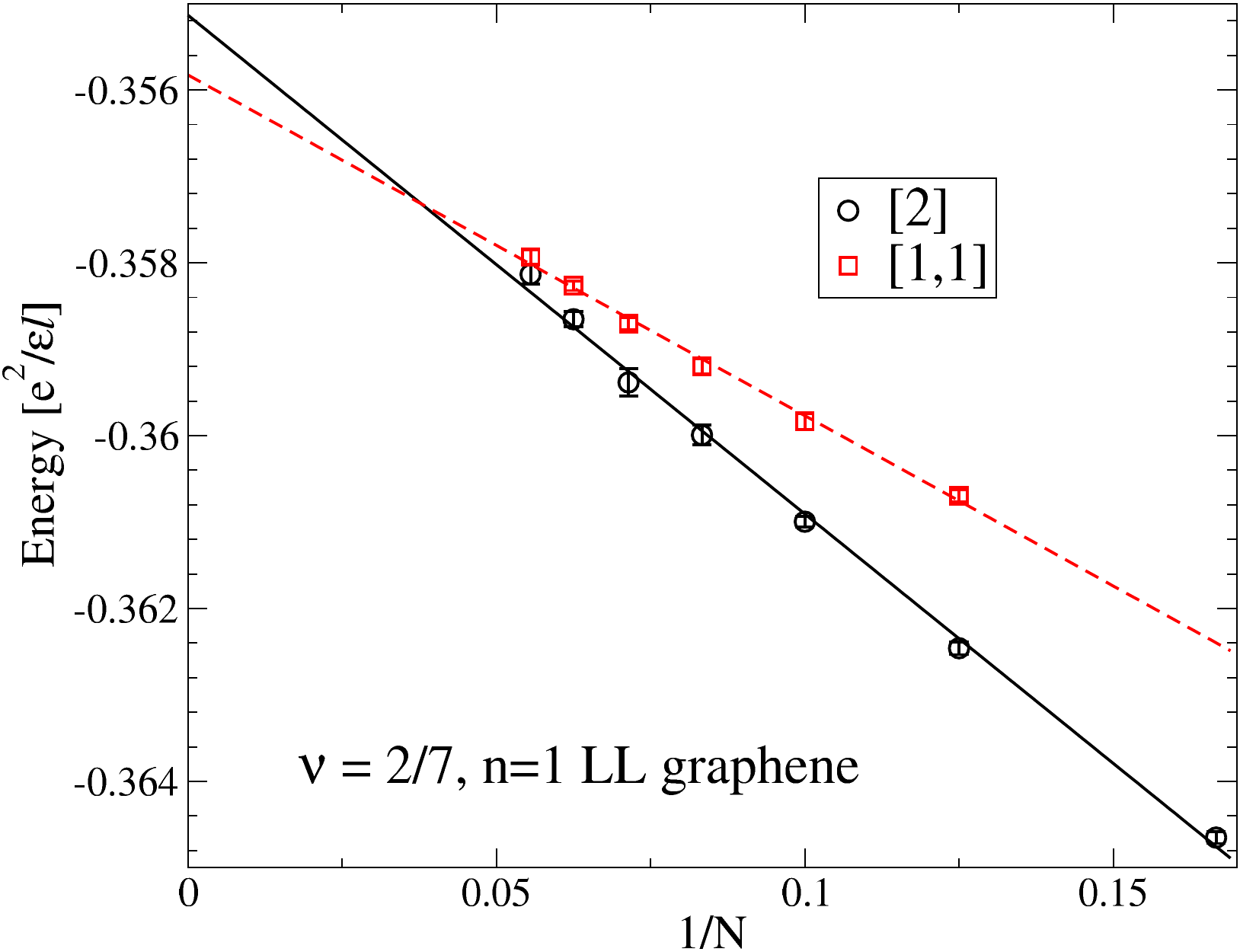}
\includegraphics[width=0.64\columnwidth,keepaspectratio]{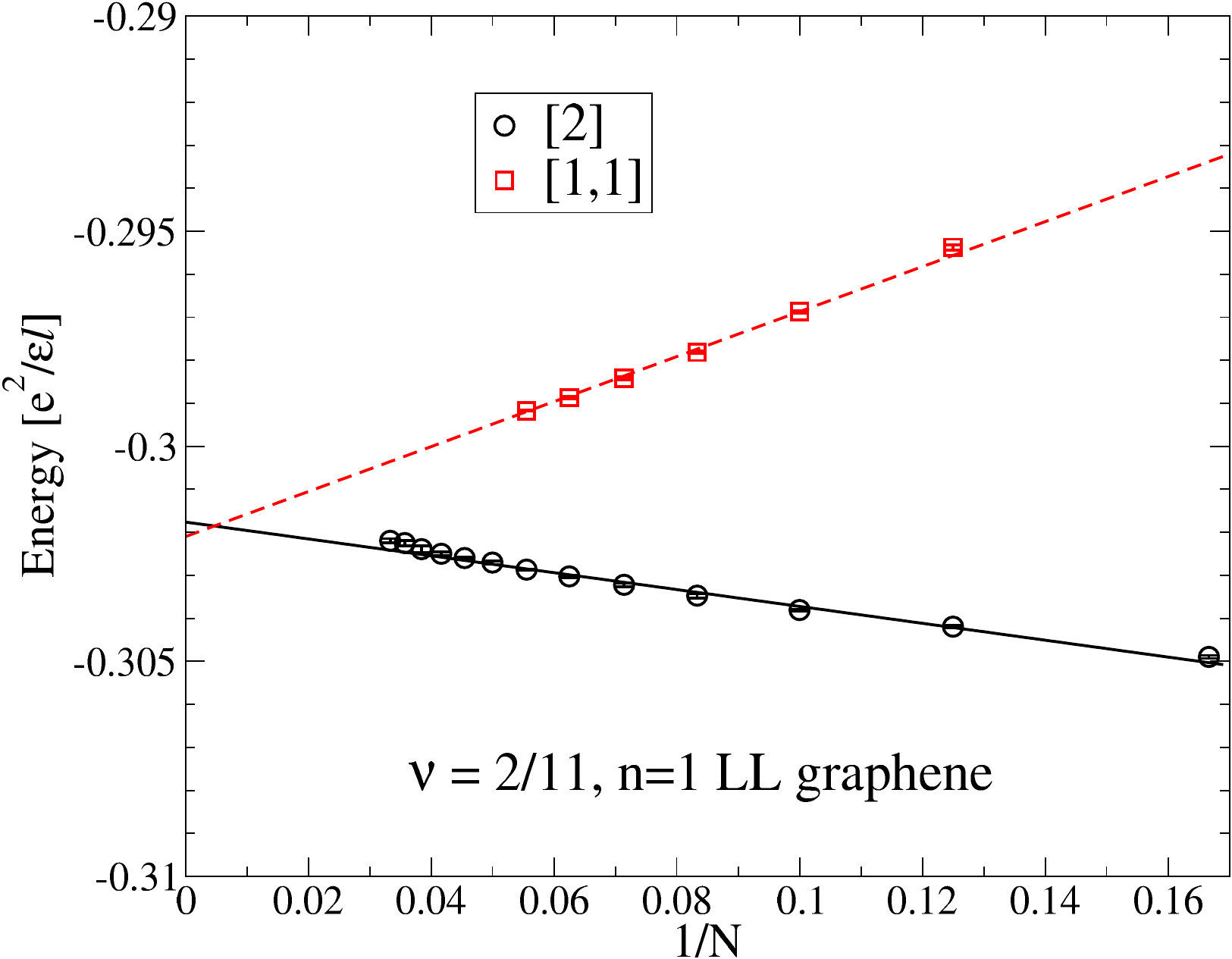}

\includegraphics[width=0.64\columnwidth,keepaspectratio]{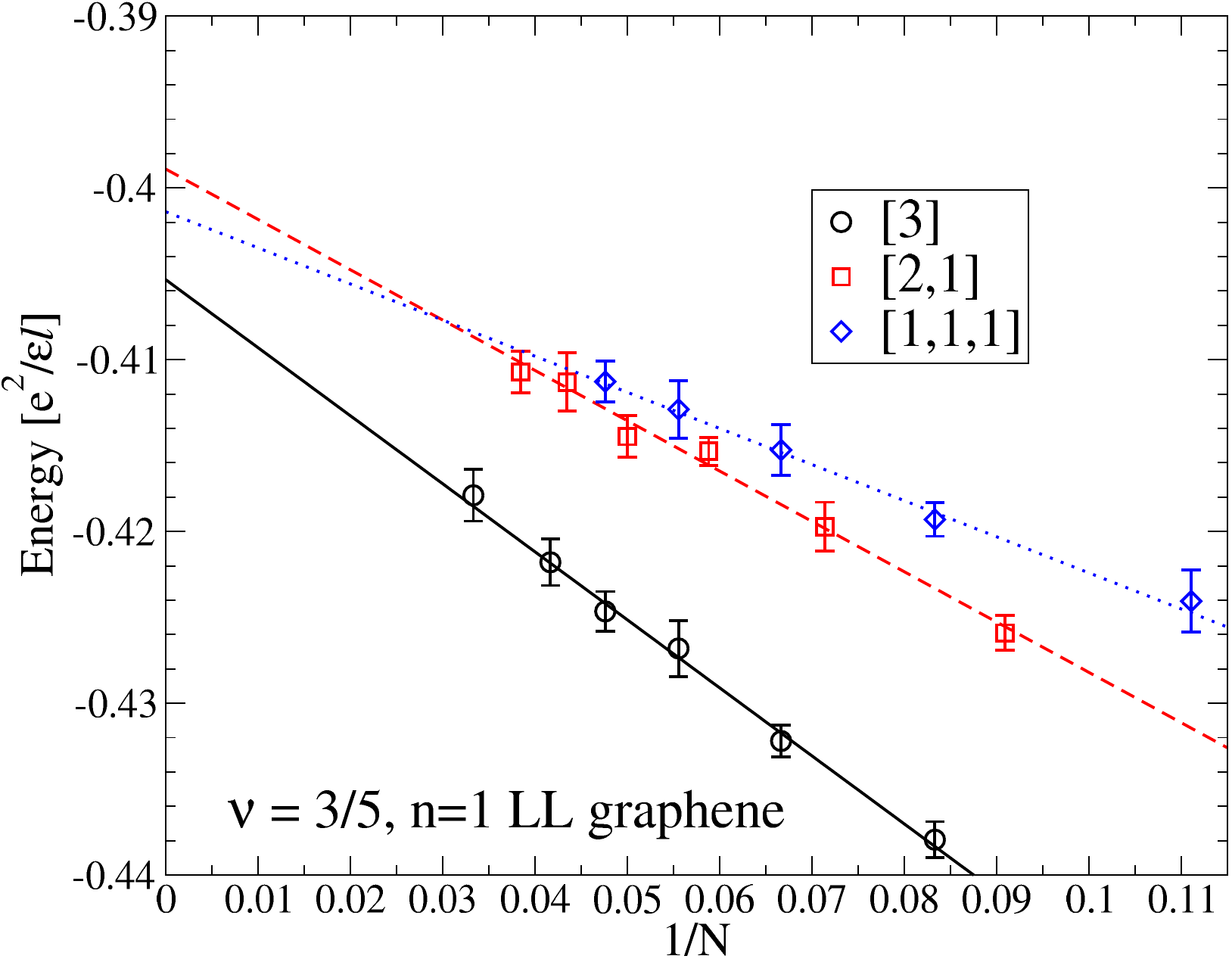}
\includegraphics[width=0.64\columnwidth,keepaspectratio]{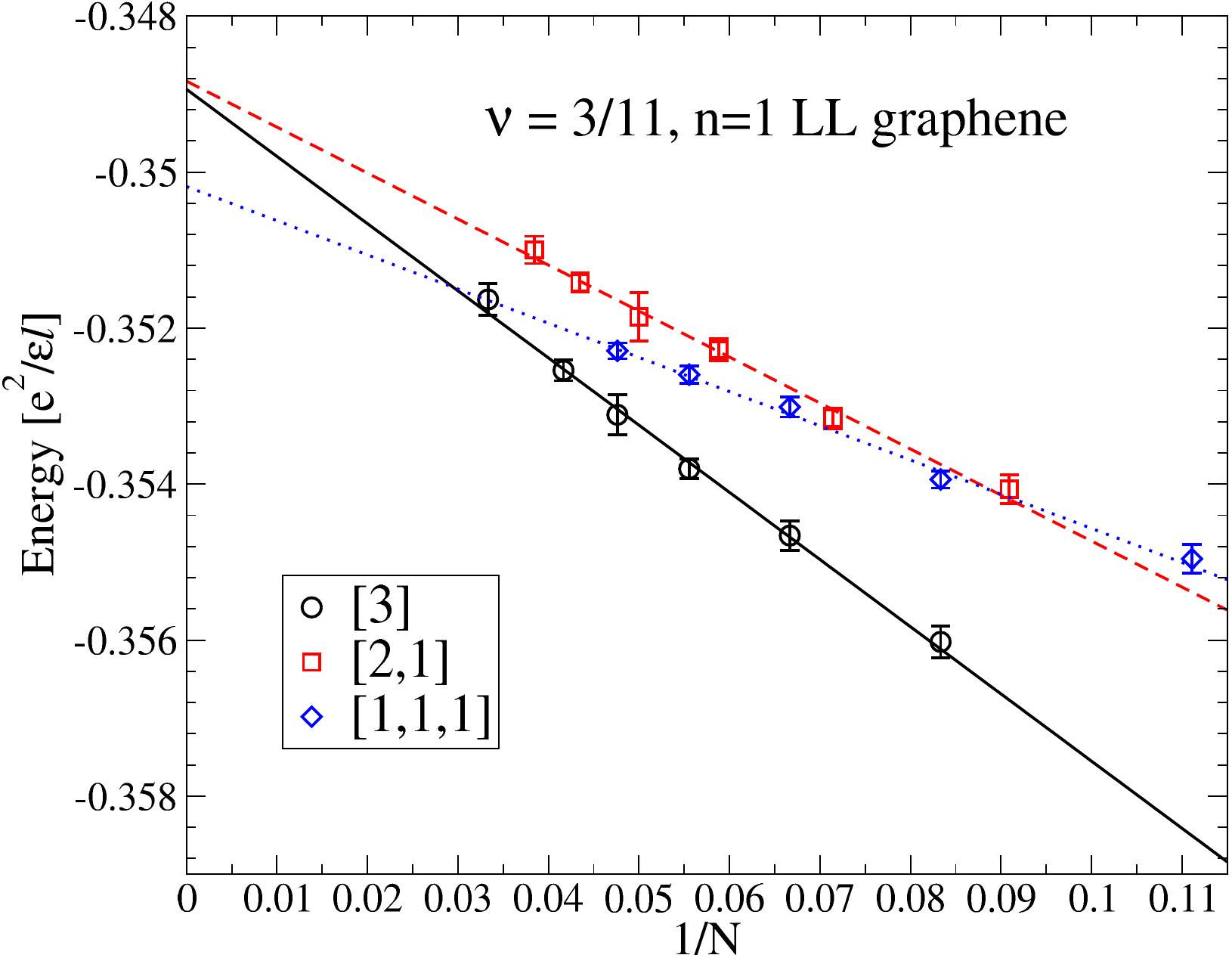}
\includegraphics[width=0.64\columnwidth,keepaspectratio]{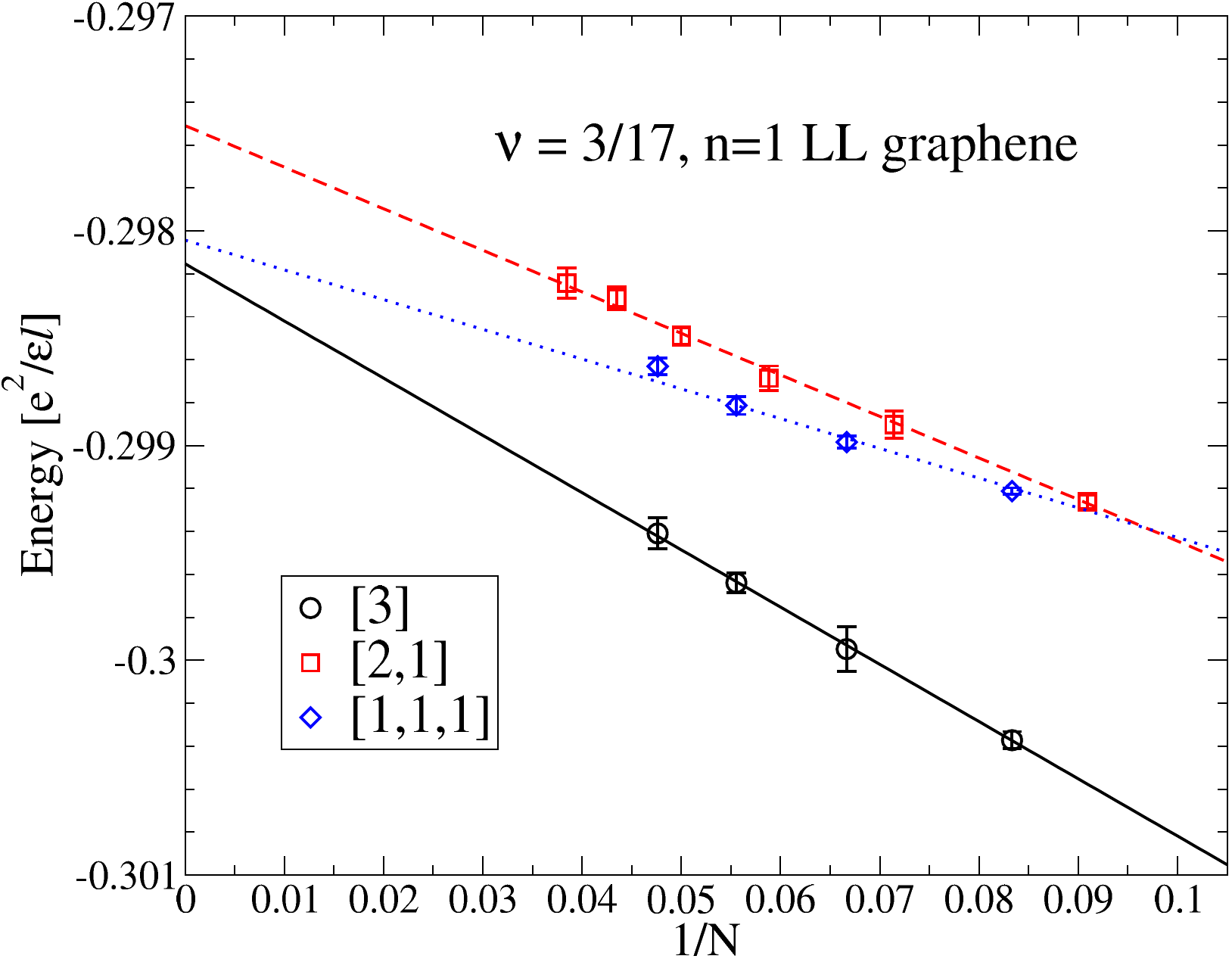}
\end{center}
\caption{Extrapolation of the energy of the wave function in Eq.~\ref{Jain_wf} to the thermodynamic limit in the $n=1$ Landau level of graphene for FQHE states in the sequence $\nu=s/(2ps-1)$.
We show the results of both linear and quadratic fits.}
\label{extra1}
\end{figure*}

\begin{figure*}[htbp]
\begin{center}
\includegraphics[width=0.64\columnwidth,keepaspectratio]{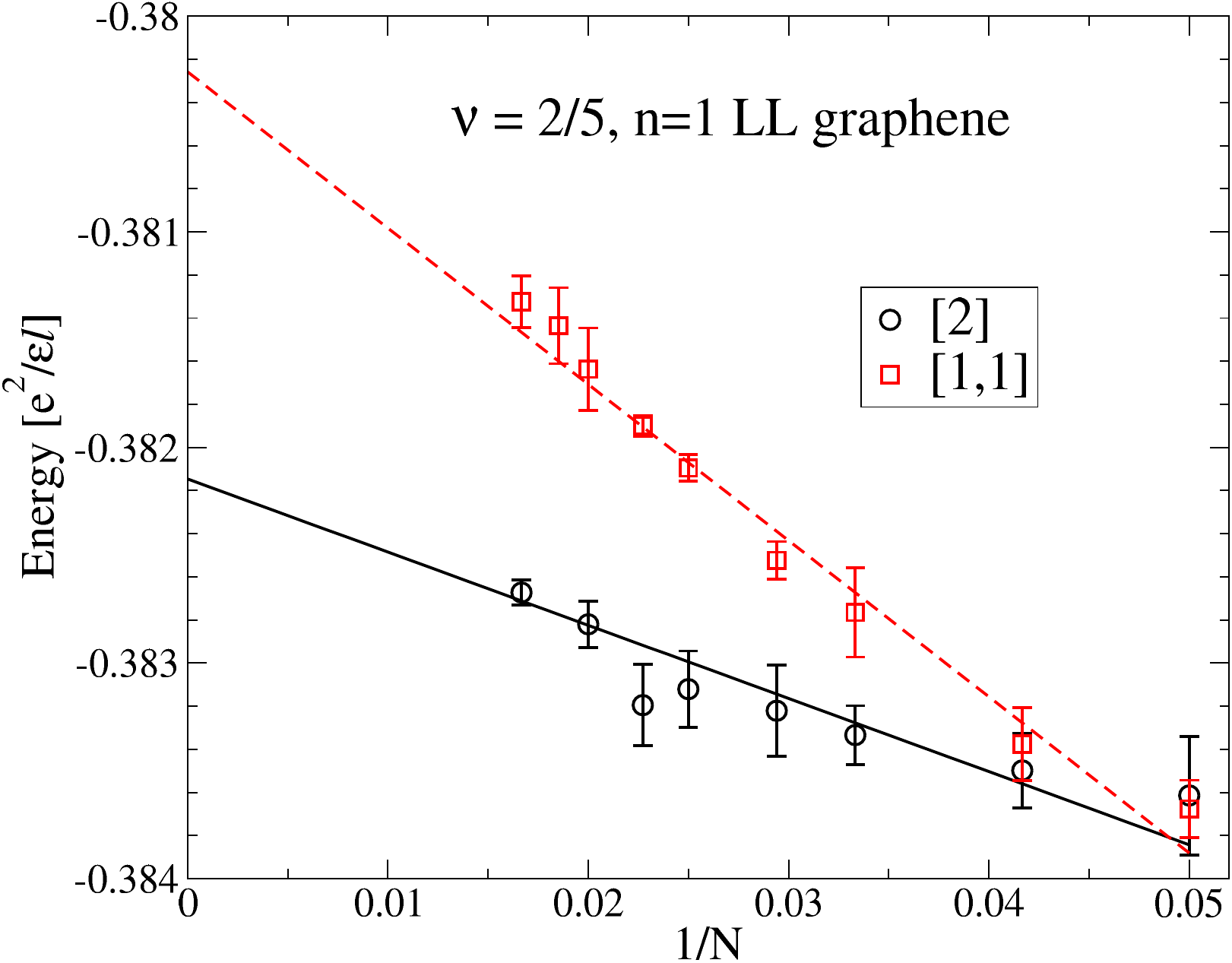}
\includegraphics[width=0.64\columnwidth,keepaspectratio]{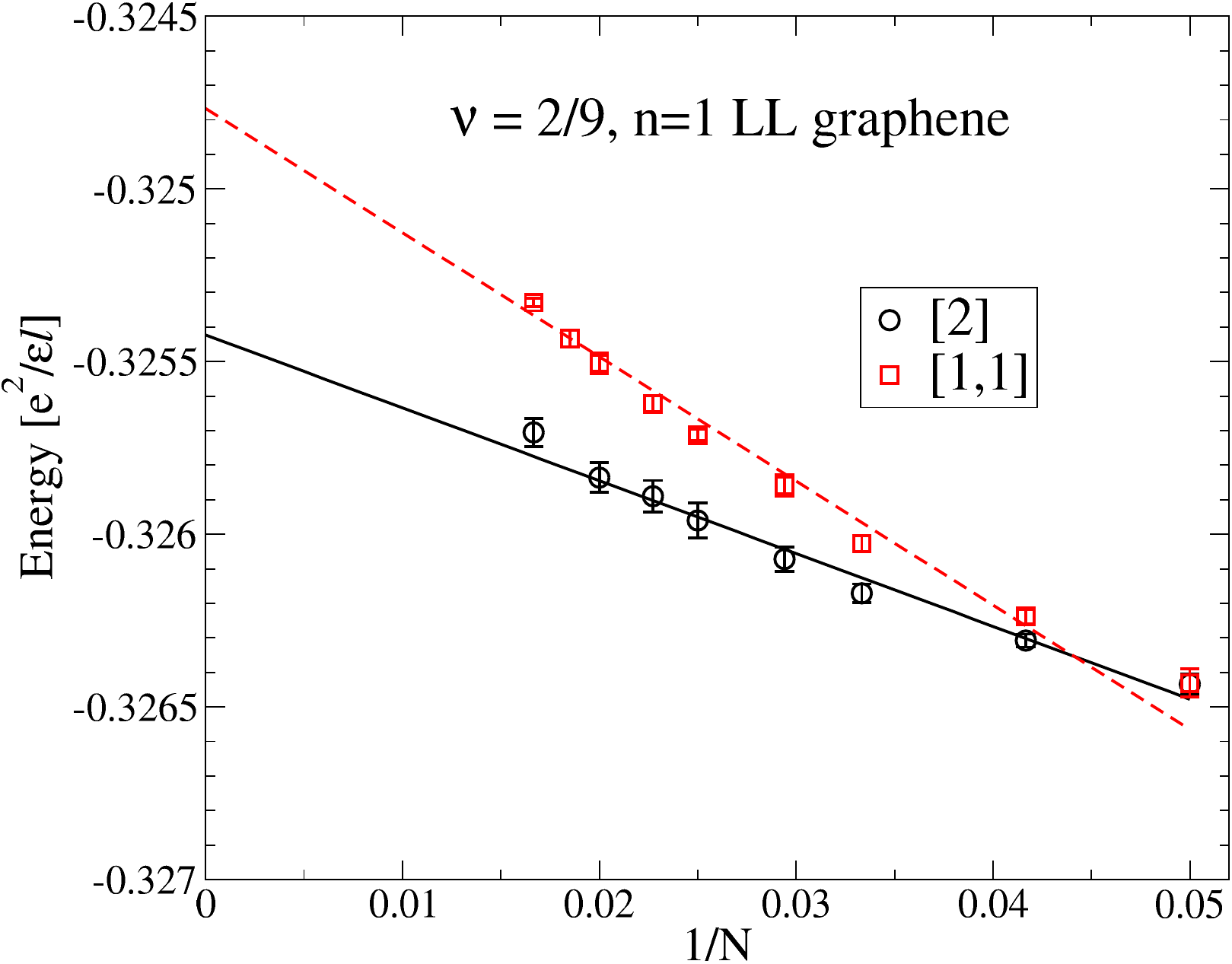}
\includegraphics[width=0.64\columnwidth,keepaspectratio]{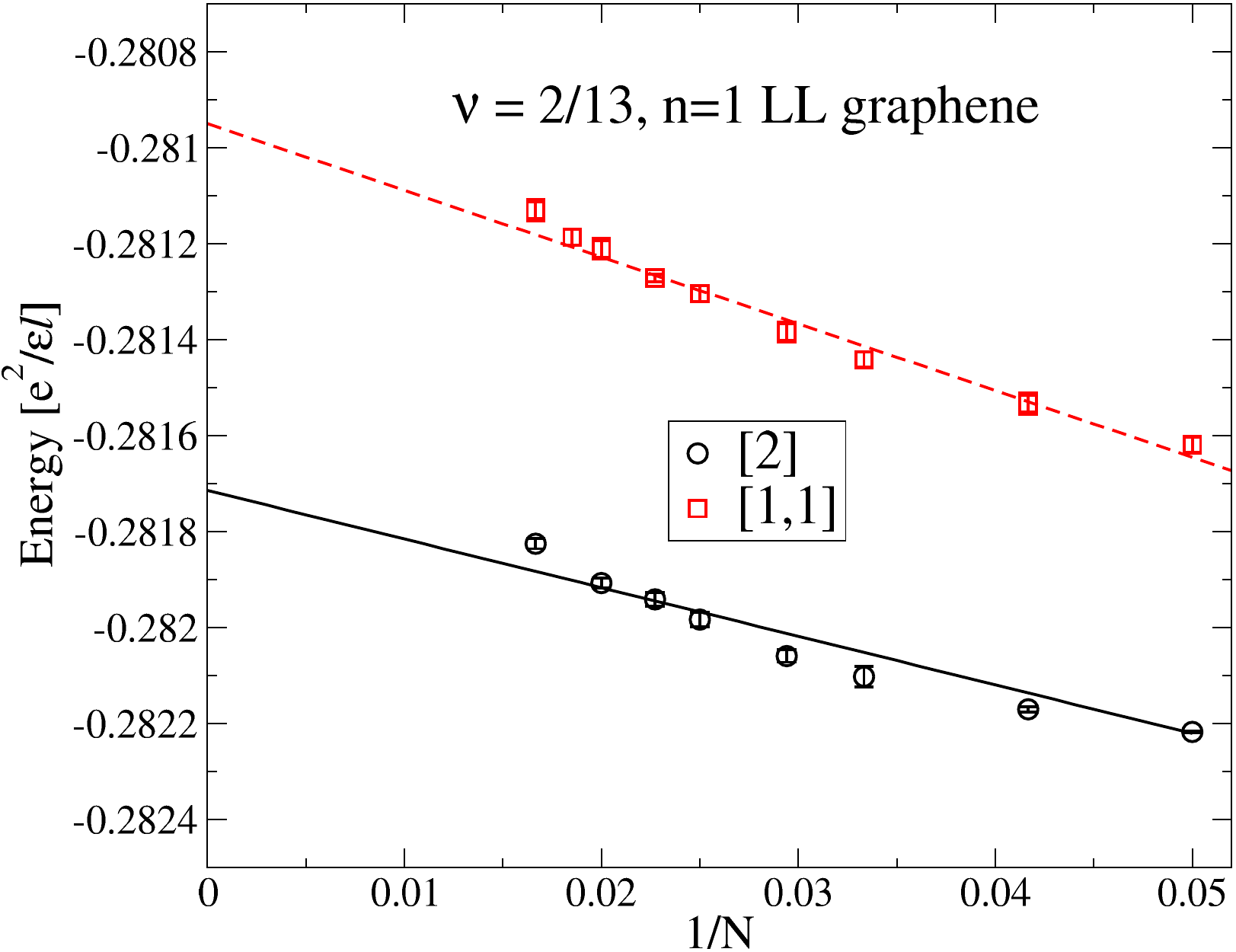}

\includegraphics[width=0.64\columnwidth,keepaspectratio]{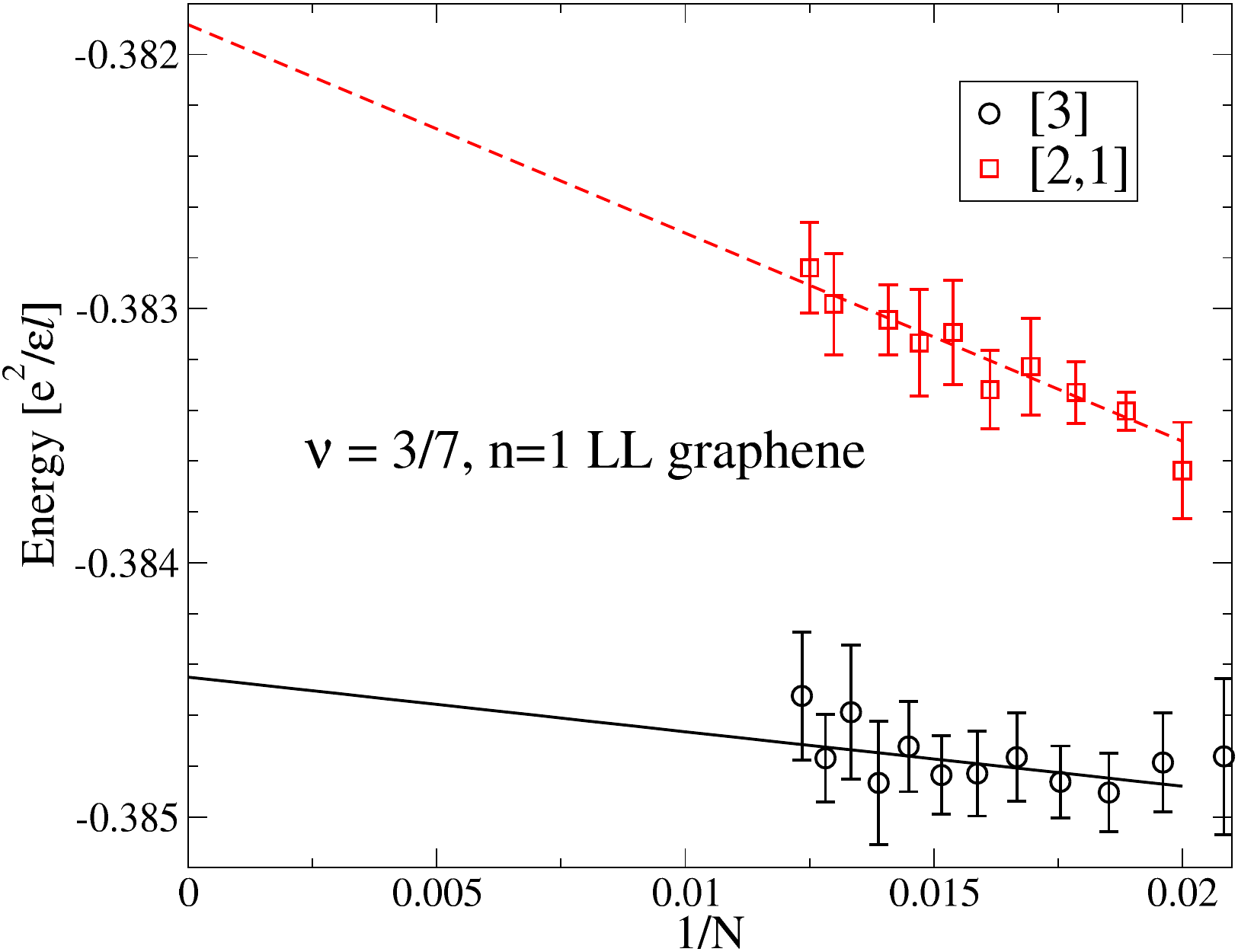}
\includegraphics[width=0.64\columnwidth,keepaspectratio]{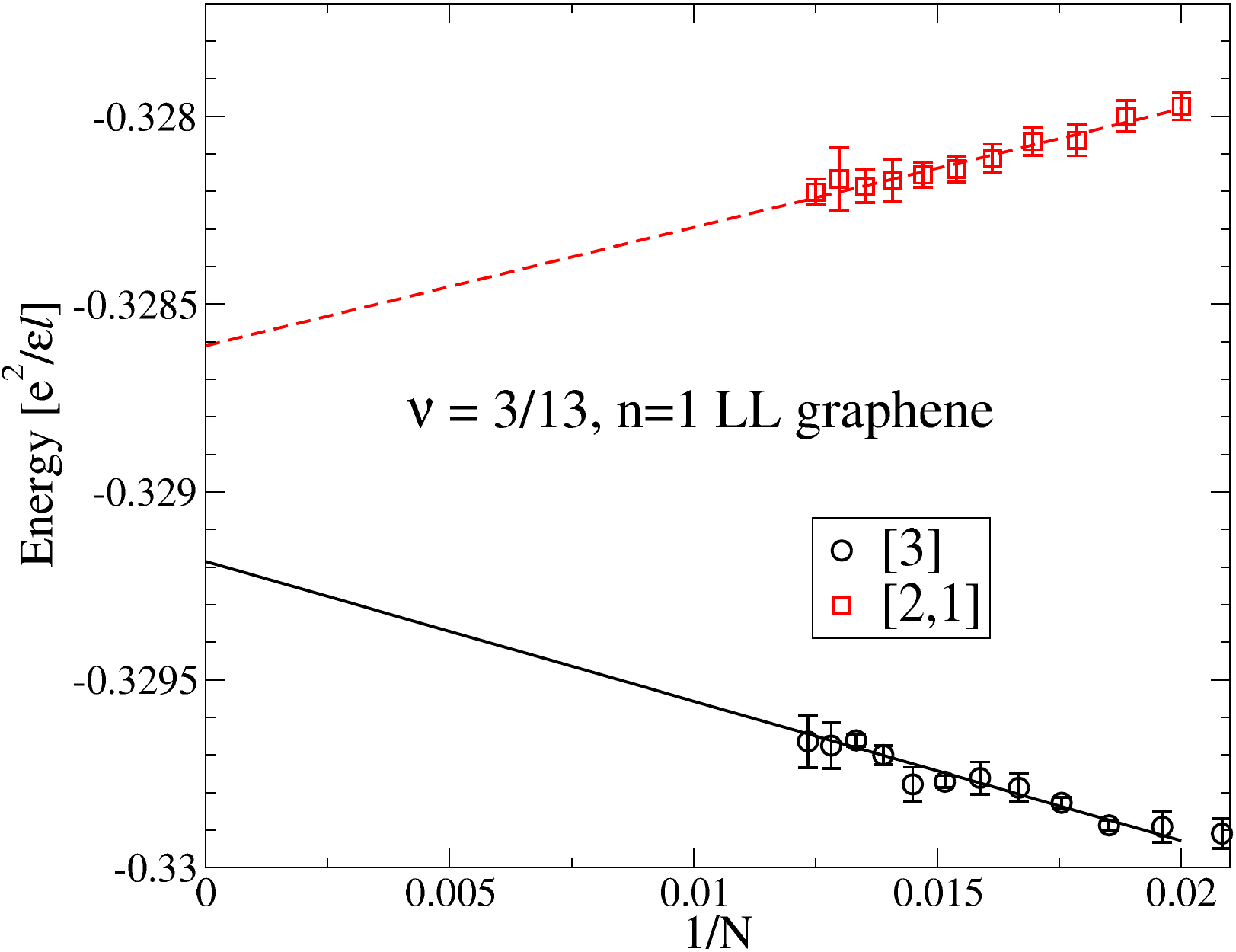}
\end{center}
\caption{Extrapolation of the energy of the wave function in Eq.~\ref{Jain_wf} to the thermodynamic limit in the $n=1$ Landau level of graphene for states in the sequence $\nu=s/(2ps+1)$.
We show the results of both linear and quadratic fits.}
\label{extra2}
\end{figure*}

The ground state energies obtained from CF wave functions of Eq. \ref{Jain_wf} by linear extrapolation to the thermodynamic limit of the different spin polarized states at various filling factors are shown in Tables \ref{tab:grn_2} and \ref{tab:grn_3}. Figs.~\ref{extra1} and \ref{extra2} show both the linear and the quadratic extrapolations of the ground state energies. We note that the results obtained by the two procedures are consistent, except perhaps at $\nu=3/17$ where the number of data points is too small for a definitive conclusion. In most cases, the energy difference is hardly affected by the extrapolation method, indicating that the slight curvature of the data points has no effect on the thermodynamic limit. We believe that this curvature is related to the background subtraction method, which does know about the effective $n=1$ LL interaction.

\begin{figure}
\begin{center}
\includegraphics[width=0.5\textwidth]{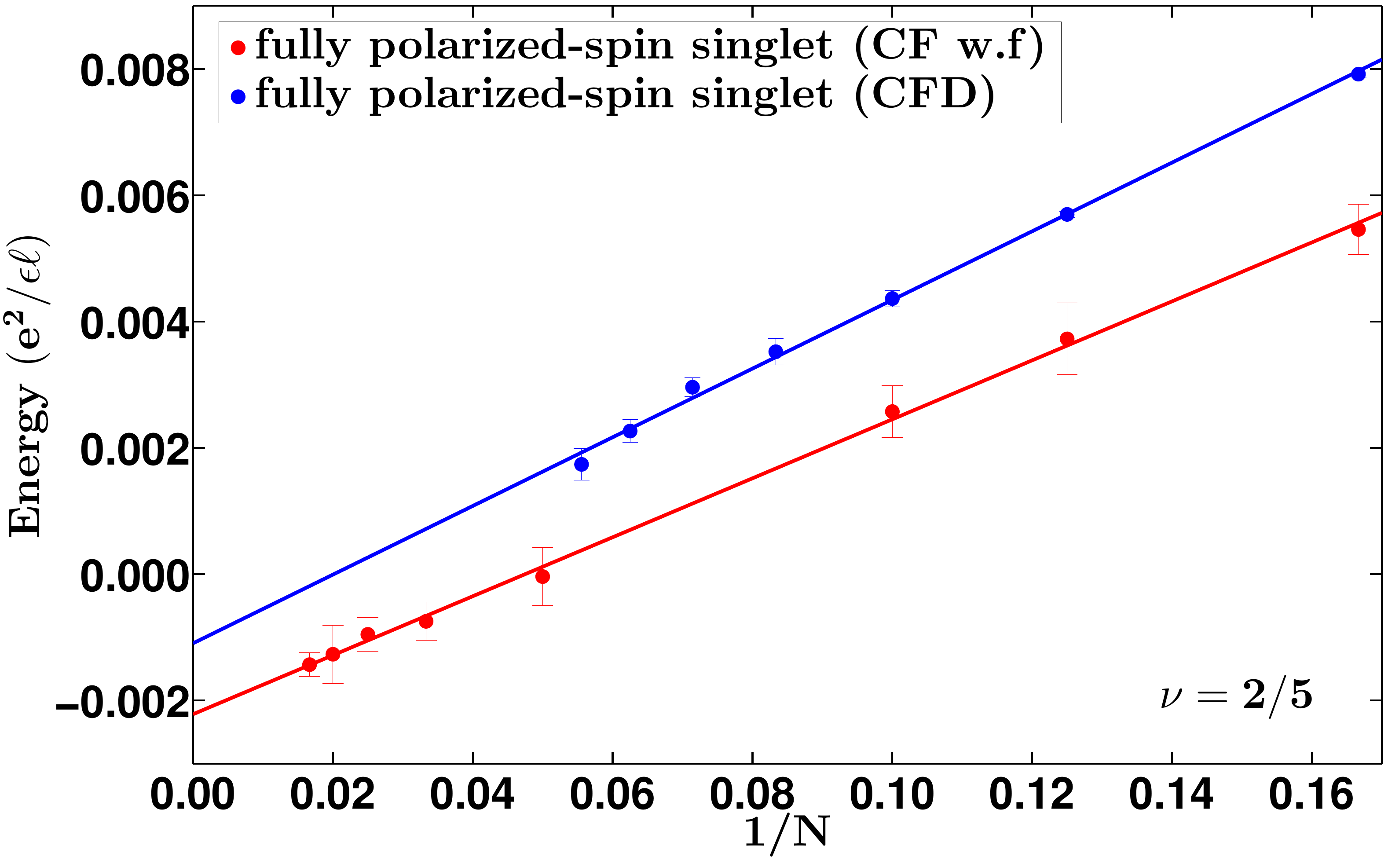}
\caption{(color online) Extrapolation to the thermodynamic limit for the difference in the fully polarized and spin-singlet $\nu=2/5$ states in the $n=1$ graphene LL. The red symbols show the energies obtained from the wave functions in Eq.~\ref{Jain_wf}, while the blue symbols are obtained from CFD.  The energies are obtained using the effective interaction of Eq. \ref{eq:eff_int}. The ground state is seen to be fully polarized.}
\label{fig:2_5}
\end{center}
\end{figure}

We find that for all states in the sequence $\nu=s/(2s\pm1)$, the fully polarized state has lower energy than the unpolarized state. In Fig. \ref{fig:2_5} we plot the difference in the energies between the fully polarized and spin-singlet states at 2/5 as a function of $1/N$. In the thermodynamic limit the fully polarized state is lower in energy than the spin-singlet state by $\sim0.002e^2/\epsilon\ell$. As noted above, the wave functions of Eq. \ref{Jain_wf} are very accurate for the fully spin polarized states but less so for non-fully spin polarized states, and one may ask if the energy ordering may change in a more accurate calculation.  In the next subsection, we present more accurate results from CF diagonalization, which show that the wave functions of Eq. \ref{Jain_wf} overestimate the difference in energy between the fully polarized and partially polarized states but still produce the correct ordering.

Next we discuss states of composite fermions carrying four and six vortices. Here again we find that the partially polarized states have higher energy as compared to the fully polarized, with a sole exception at $\nu=2/7$, where we find that the spin-singlet state has lower energy in comparison to the fully polarized state. We predict the the critical Zeeman energy for the transition from a spin-singlet state to a fully polarized state occurs at a Zeeman energy of $0.0014(4)~e^2/\epsilon\ell$. We note that in the $n=0$ GLL too, for composite fermions carrying more than two vortices, $2/7$ is the only filling factor in which the ground state is unpolarized \cite{Balram15a}. As noted before, our calculation is inconclusive at $\nu=3/17$; due to its moderate experimental relevance we do not pursue this issue further.

\begin{figure}
\begin{center}
\includegraphics[width=0.5\textwidth]{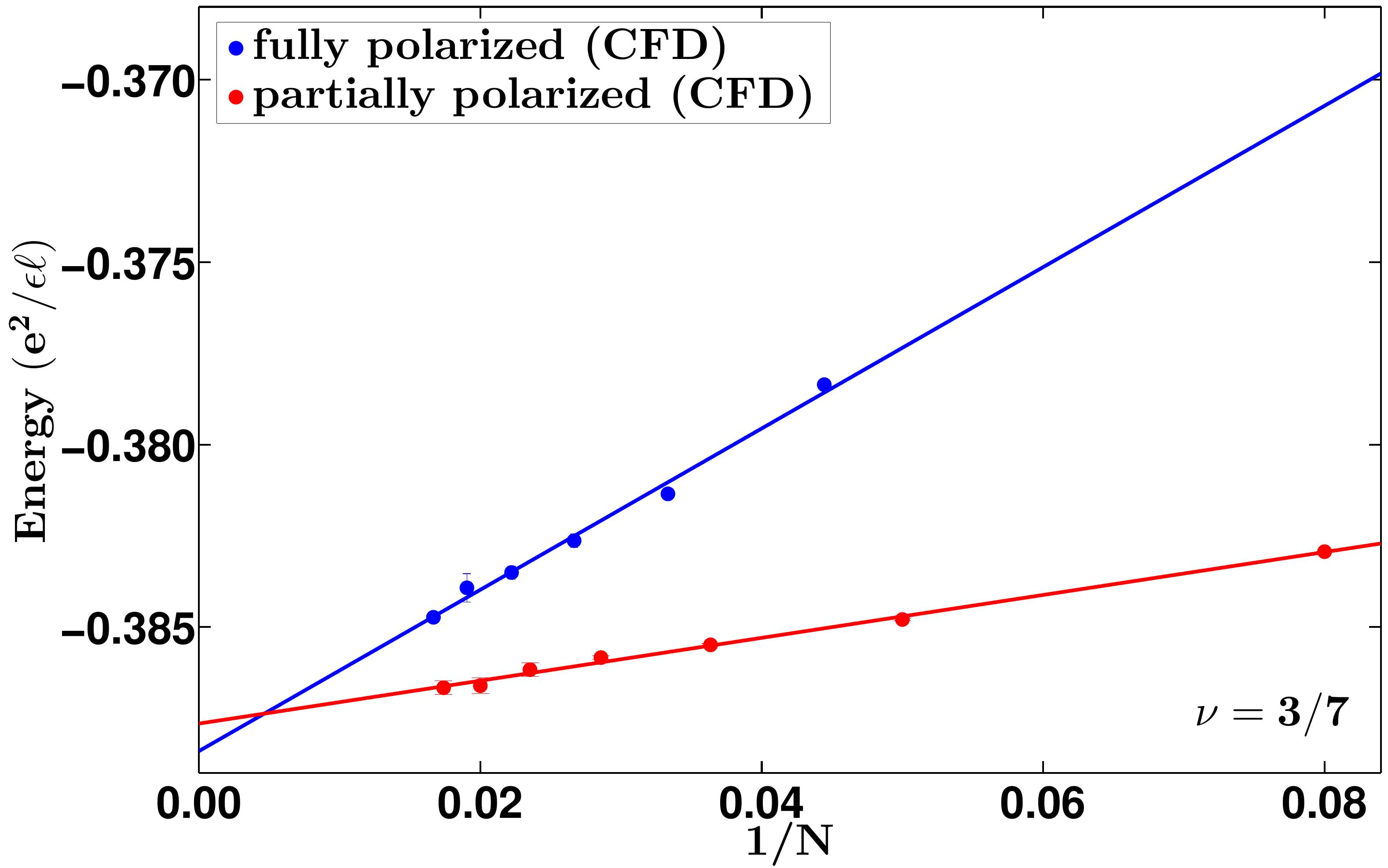}
\caption{(color online) Extrapolation to the thermodynamic limit for the fully polarized (blue) and partially polarized (red) $\nu=3/7$ CFD energies for the $n=1$ GLL obtained using the effective interaction of Eq. \ref{eq:eff_int}. The ground state is seen to be fully polarized.}
\label{fig:3_7}
\end{center}
\end{figure}

\subsection{CF diagonalization}
We have carried out the procedure of CFD to calculate the the ground state energies of the fully polarized and unpolarized states at two filling factors: 2/5 and 3/7. The linear extrapolation of the fully polarized and unpolarized ground state energies obtained from CFD at 3/7 are shown in Fig. \ref{fig:3_7}. The fully polarized and spin-singlet ground state energies are not well approximately by a linear function of 1/N for 2/5, but the difference in the two energies is. In Fig. \ref{fig:2_5} we show this difference for 2/5. In the thermodynamic limit we find that the fully polarized state has lower energy than the spin-singlet state. For $3/7$ the thermodynamic energy obtained from CFD for the fully polarized and partially polarized states are -0.3884(3) and -0.3877(1) respectively. As with 2/5, we again find that the fully polarized state has lower energy than the partially polarized state. The CFD results show, as anticipated, a smaller difference of $\sim0.001~e^2/\epsilon\ell$ between the two energies than that obtained from the CF wave functions of Eq. \ref{Jain_wf}.

\begin{table}
\begin{center}
\begin{tabular}{|c|c|c|}
\hline
\multicolumn{1}{|c|}{$\nu$} & \multicolumn{1}{|c|}{fully polarized (CF w. f.)} & \multicolumn{1}{|c|}{spin singlet (CF w. f.)}\\ \hline
2/3 	    & -0.4122(4)	& -0.4097(7)   \\ \hline
2/5 	    & -0.3821(1) 	& -0.3803(1)   \\ \hline
2/7 	    & -0.3551(1) 	& -0.3558(1)   \\ \hline
2/9 	    & -0.32542(4) 	& -0.32477(2)  \\ \hline
2/11 	    & -0.30176(1) 	& -0.30170(3)  \\ \hline
2/13 	    & -0.281714(9) 	& -0.28095(2)  \\ \hline
\end{tabular}
\end{center}
\caption {Thermodynamic limit of the $n=1$ GLL energies obtained using the effective interaction of Eq. \ref{eq:eff_int} for the fully polarized and spin singlet states at $\nu=2/(4p\pm1)$. All energies are in units of $e^{2}/\epsilon\ell$. These numbers are obtained using the CF wave functions in Eq.~\ref{Jain_wf}. Only for 2/7 does the spin singlet state have a lower energy than the fully polarized state.} 
\label{tab:grn_2}
\end{table}

\begin{table}
\begin{center}
\begin{tabular}{|c|c|c|}
\hline
\multicolumn{1}{|c|}{$\nu$} & \multicolumn{1}{|c|}{fully polarized (CF w. f.)} & \multicolumn{1}{|c|}{partially polarized (CF w. f.)}\\ \hline
3/5 	    & -0.405(2)	    & -0.399(2)	   \\ \hline
3/7 	    & -0.3845(4)    & -0.3819(3)   \\ \hline
3/11 	    & -0.3489(3)    & -0.3488(2)   \\ \hline
3/13 	    & -0.32919(5)   & -0.32861(8)  \\ \hline
3/17 	    & -0.29803(9)   & -0.29751(7)  \\ \hline
\end{tabular}
\end{center}
\caption {Same as in Table \ref{tab:grn_2} but for $\nu=3/(6p\pm1)$. We find that the fully polarized state has lower energy than the partially polarized one for each filling factor.} 
\label{tab:grn_3} 
\end{table}

\subsection{Corrections from Landau level mixing}
The strength of LL mixing is quantified by a parameter $\lambda$, which is the ratio of the interaction to the kinetic energy\cite{Peterson13}. In graphene this parameter is given by $\lambda=(e^2/\epsilon \ell)/(\hbar v_{\rm F}/\ell)=e^2/(\hbar\epsilon v_{\rm F})$, which also equals the graphene fine-structure constant. Unlike in the case of parabolic bands (as would be appropriate to semiconductor quantum wells such as GaAs), $\lambda$ is independent of the external magnetic field for electrons with a linear dispersion. For different substrates the values of $\lambda$ are as follows: $\lambda\approx2.2$ for suspended graphene, $\lambda\approx0.9$ for graphene on silicon dioxide, $\lambda=0.5-0.8$ for graphene on boron nitride\cite{DasSarma11,Peterson13}. 

We note that unlike in the $n=0$ GLL, in $n=1$ GLL, LL mixing does produce an effective three body interaction in addition to the shifts to the pairwise interaction\cite{Peterson14}. We have calculated the relevant (short range) pair and triplet amplitudes per electron in the considered  ground states known for a series of system sizes $N$, then extrapolated these amplitudes to an infinite system, and finally convoluted the extrapolated amplitudes with pair and triplet pseudopotentials of the effective interaction which accounts for Landau level mixing\cite{Peterson13}.

\begin{figure}
\begin{center}
\includegraphics[width=0.5\textwidth,height=0.28\textwidth]{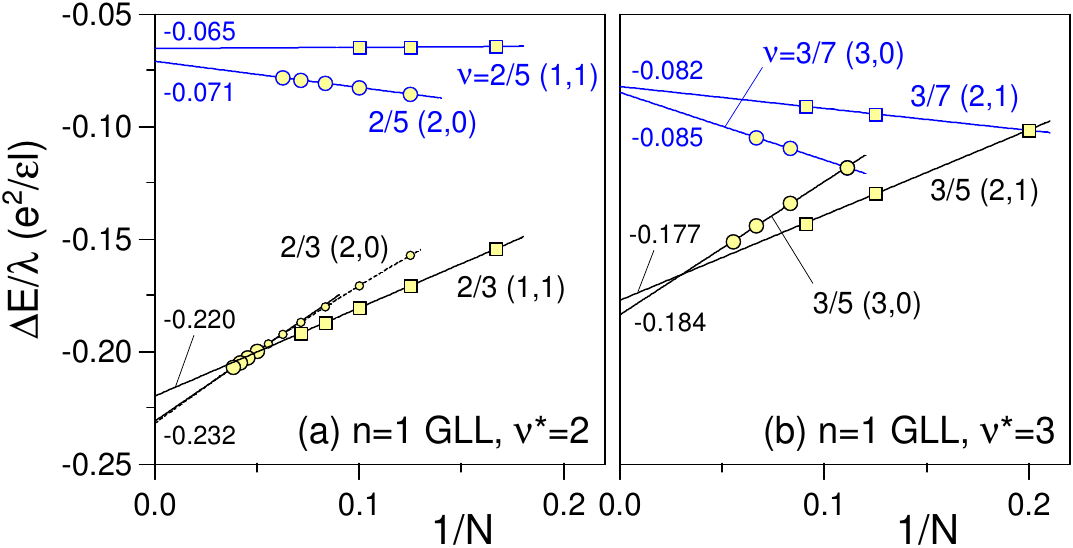}
\caption{(color online) Thermodynamic extrapolation of the exact Coulomb ground state energies per unit LL mixing parameter $\lambda$ for the fully polarized (cirlces) and unpolarized states (squares) in the $n=1$ GLL at filling factors 2/3 and 2/5 (a) and 3/5 and 3/7 (b).}
\label{fig:exact_gs_LLmix}
\end{center}
\end{figure}

Fig. \ref{fig:exact_gs_LLmix} shows the energy shifts induced by LL mixing for the fully polarized and unpolarized ground states in $n=1$ GLL. As is evident from the figure, LL mixing favors fully polarized states over unpolarized states. We note that Ref.~\onlinecite{Peterson14} does not give the $3$-body pseudopotential for spin $s=1/2$ and $m=3$, which we arbitrarily set to zero. However, the corresponding amplitudes are very small, so we do not expect that adding the actual value of $V(s=1/2,m=3)$ would affect the result. We end this subsection with the caveat that the corrections in Ref.~\onlinecite{Peterson14} are obtained within a perturbative scheme, and we have not tested to what extent this approach remains applicable for the realistic values of the LL mixing parameter $\lambda$ (See Ref. \onlinecite{Balram15a} for a detailed discussion). 

\section{Excitations}
\label{sec:ex_studies}
The FQHE ground state at $\nu=s/(2ps\pm1)$ are explained as $s$ $\Lambda$Ls of composite fermions carrying $2p$ vortices. Furthermore, their low-energy excitations are obtained by exciting composite fermions across $\Lambda$Ls \cite{Dev92,Jain07,Balram13}.  Our motivation for investigating the excitations comes from the fact that spin reversed excitations of composite fermions can reveal an instability of the fully polarized ground state at sufficiently small Zeeman energies. Such an instability is signaled by the presence of a sub-Zeeman energy roton in the dispersion of the spin reversed neutral exciton\cite{Mandal01,Wurstbauer11}; theoretically, it is necessary to dress the single spin reversed CF exciton by other excitations to capture the physics of the spin roton.

We have studied two kinds of low-energy collective modes at the filling factors $\nu=s/(2s\pm1)$ assuming the ground state to be fully polarized. These are: \\
(i) \underline{Spin-conserving}: a CF is excited from the $(s-1)$$\uparrow\Lambda$L to the $s$$\uparrow\Lambda$L. These excitons form the \emph{magnetoroton/magnetoplasmon} mode. Their energies have been calculated theoretically for many of the fractions in the LLL\cite{Jain98}. Resonant inelastic light scattering experiments have observed these modes and their measured energies are consistent with the theoretical predictions \cite{Pinczuk93,Scarola00,Groshaus08}. \\
(ii) \underline{Spin-reversing}: a CF is excited from the $(s-1)$$\uparrow\Lambda$L to the $0$$\downarrow\Lambda$L. These excitons form the \emph{spin-flip} mode. The energy of these modes are evaluated numerically in the lowest LL and it was found that for $n\geq2$ this excitation has a lower energy than the ground state up to a finite Zeeman energy showing that the fully polarized ground state is unstable to these excitations \cite{Mandal01}. This is consistent with the fact the ground state in the LLL at zero Zeeman energies is spin unpolarized\cite{Park98,Balram15a}. For comparison we have also evaluated the energy of the spin-wave mode. \\

\begin{figure}
\begin{center}
\includegraphics[width=0.5\textwidth]{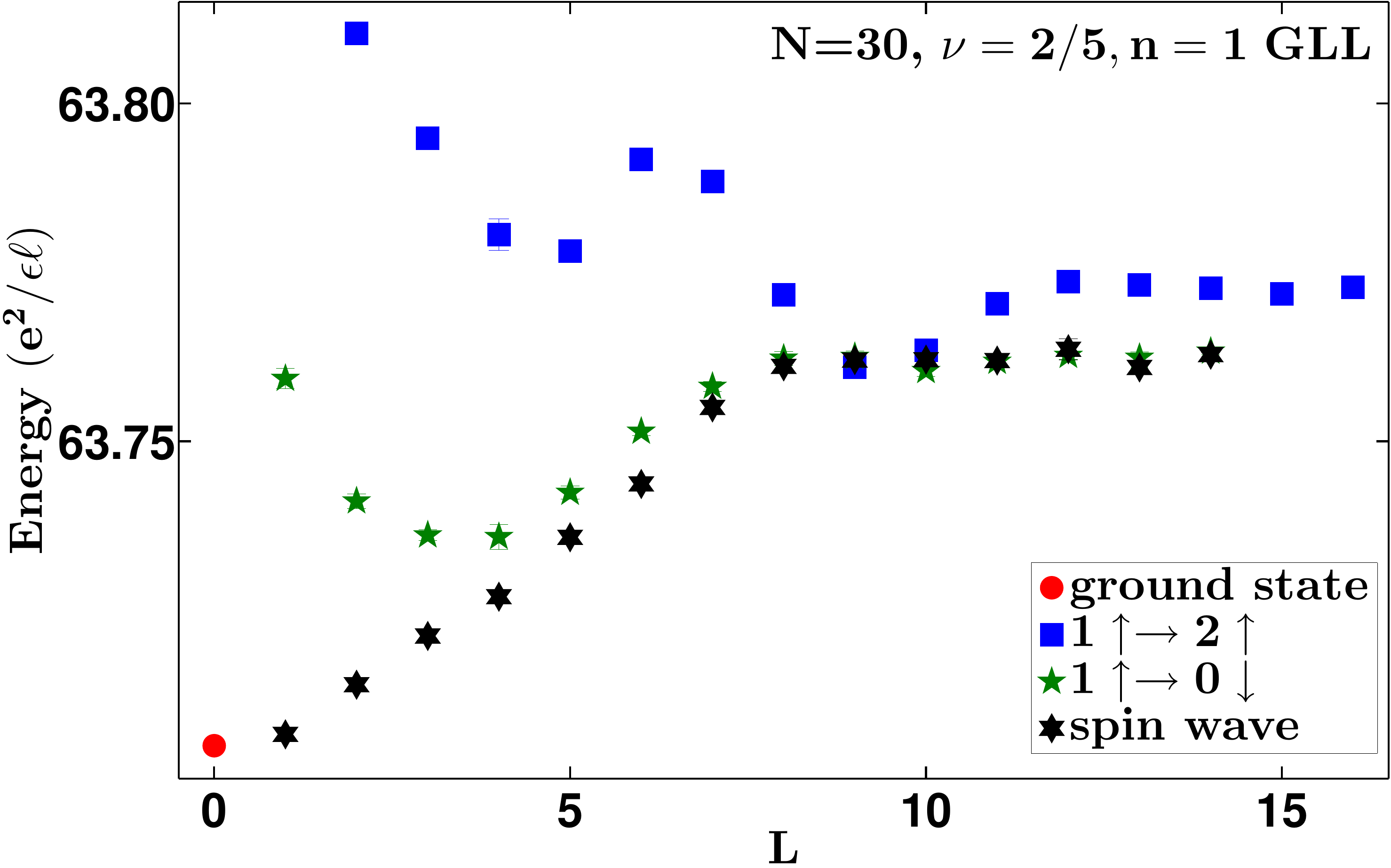}
\includegraphics[width=0.5\textwidth]{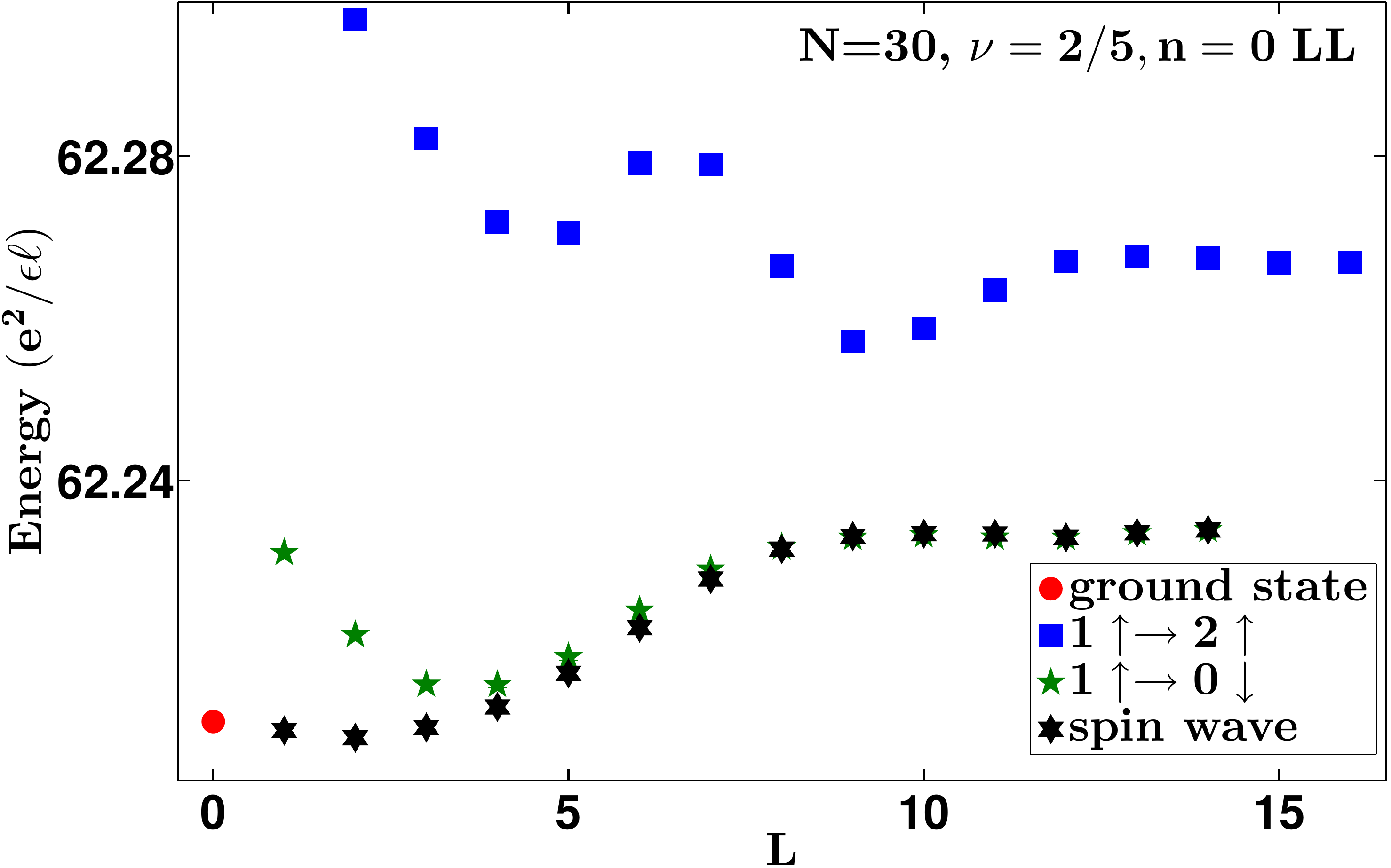}
\end{center}
\caption{(color online) Spin wave (black hexagram), spin-conserving (blue squares) and spin-reversing (green pentagram) collective mode energies for $N=30$ at $\nu=2/5$ for the $n=1$ GLL. These energies are obtained using the wave function in Eq.~\ref{Jain_wf}, and the effective interaction of Eq. \ref{eq:eff_int} (top panel). For comparison the corresponding modes in the lowest LL are shown in the bottom panel.}
\label{fig:2_5_modes}
\end{figure}

In Fig. \ref{fig:2_5_modes} we show the collective modes in the $n=1$ GLL for a fully polarized at $\nu=2/5$ obtained using the CF theory with the effective interaction of Eq. \ref{eq:eff_int}. In the bottom panel of the same figure, the corresponding modes in the lowest LL are shown for comparison. In contrast to the lowest LL \cite{Mandal01}, the spin-flip modes in the $n=1$ GLL does not have a roton minimum below its long wave length energy. Consequently, unlike the lowest LL, there is no spin-flip instability in the second GLL. It is interesting to note that in the $n=1$ GLL the large wave vector gaps for the spin conserving and spin reversing modes are almost identical, in contrast  to the LLL where the two gaps are significantly different. This is seen in the spectrum shown in Fig. \ref{fig:2_5_modes} where the two collective modes are seen to approach each other in the $n=1$ GLL in the large wave vector limit. 

\begin{figure}
\begin{center}
\includegraphics[width=0.5\textwidth]{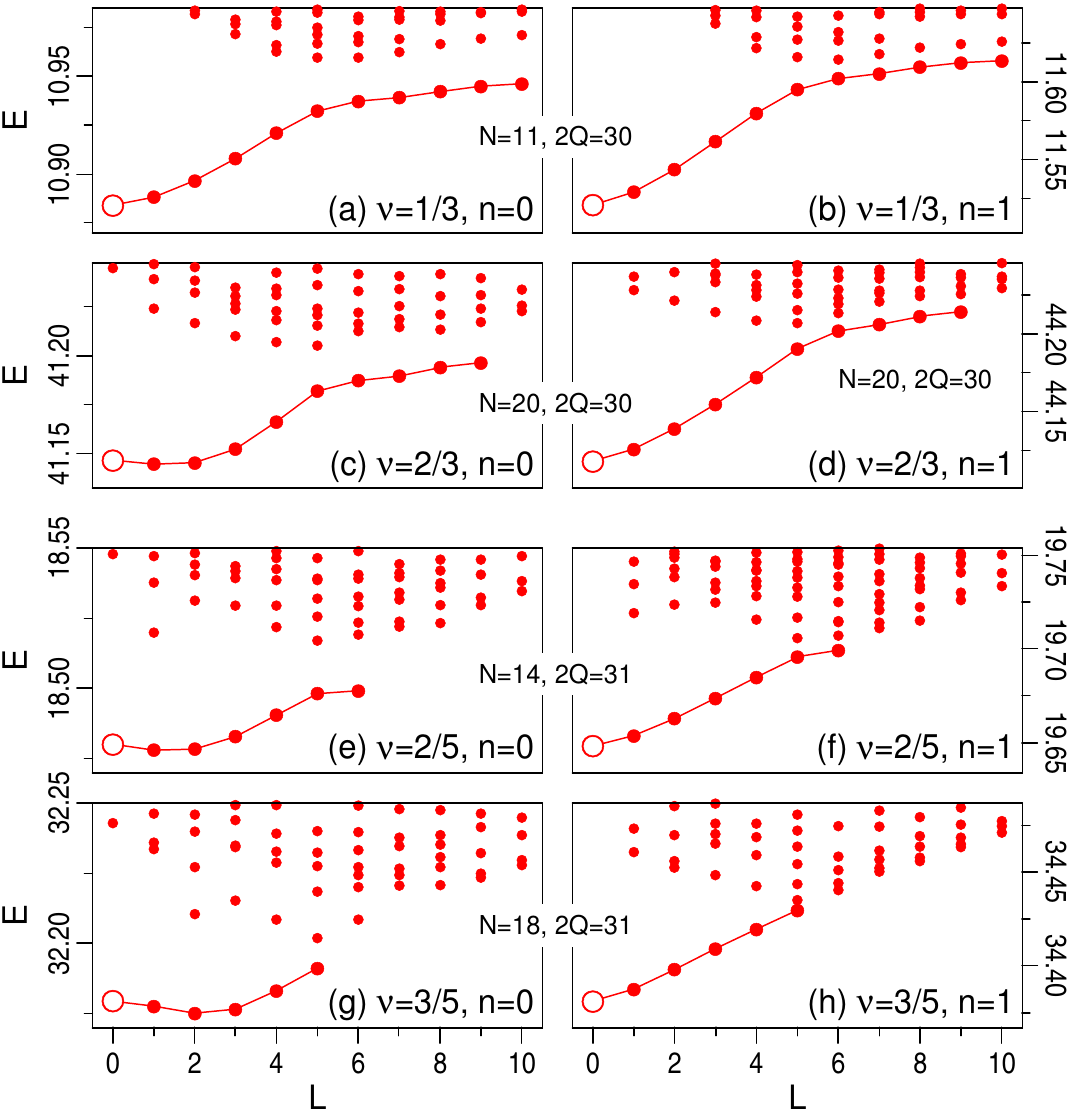}
\end{center}
\caption{(color online) Comparison of the spin wave dispersions in the LLL (left panels) and $n=1$ GLL (right panels) at various filling factors in the sequence $\nu=s/(2s\pm1)$, obtained from exact diagonalization. The spin-wave in the LLL shows a spin-flip instability for $s\geq 2$, while the $n=1$ GLL supports a robust spin wave.}  
\label{fig:exact_exc1}
\end{figure}

\begin{figure}
\begin{center}
\includegraphics[width=0.5\textwidth]{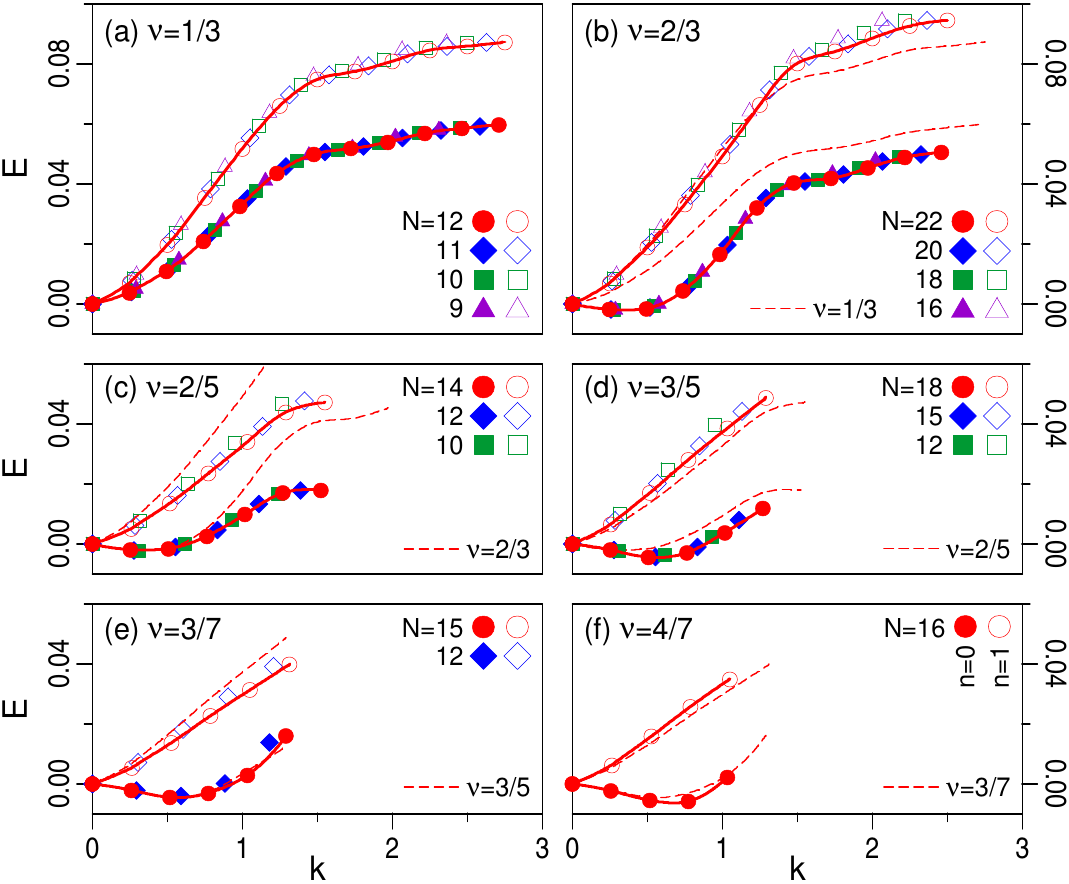}
\end{center}
\caption{(color online) Comparison of the spin wave dispersions in the LLL (filled symbols) and $n=1$ GLL (open symbols) at various filling factors in the sequence $\nu=s/(2s\pm1)$ for different system sizes. The dispersions are obtained from exact diagonalization.}
\label{fig:exact_exc2}
\end{figure}

We have also obtained the spin wave dispersion for the fully polarized state in the $n=0$ and $n=1$ GLL from exact diagonalization, shown in Figs. \ref{fig:exact_exc1} and \ref{fig:exact_exc2}. It is evident that the fully spin polarized state at filling factors $\nu=s/(2s\pm 1)$ for $s\geq 2$ in the $n=0$ GLL is unstable to a spin-flip instability at vanishing Zeeman energies whereas that in the $n=1$ GLL remains stable.

\section{Is there a pairing instability of composite fermions in the $n=1$ GLL?}
\label{sec:Pf_vs_CFFS}

Given that filling factor $\nu=1/2$ in the $n=1$ LL of GaAs produces an incompressible state, it is natural to ask if the same is true in graphene. This question has been addressed in the past using both exact diagonalization\cite{Wojs11a} and methods of CF theory\cite{Toke07b}, and it was concluded that the Pfaffian states is not favored.
For completeness, we have addressed this issue with our more accurate real-space interaction. 

\begin{figure*}[htpb]
\begin{center}
\includegraphics[width=0.64\columnwidth,keepaspectratio]{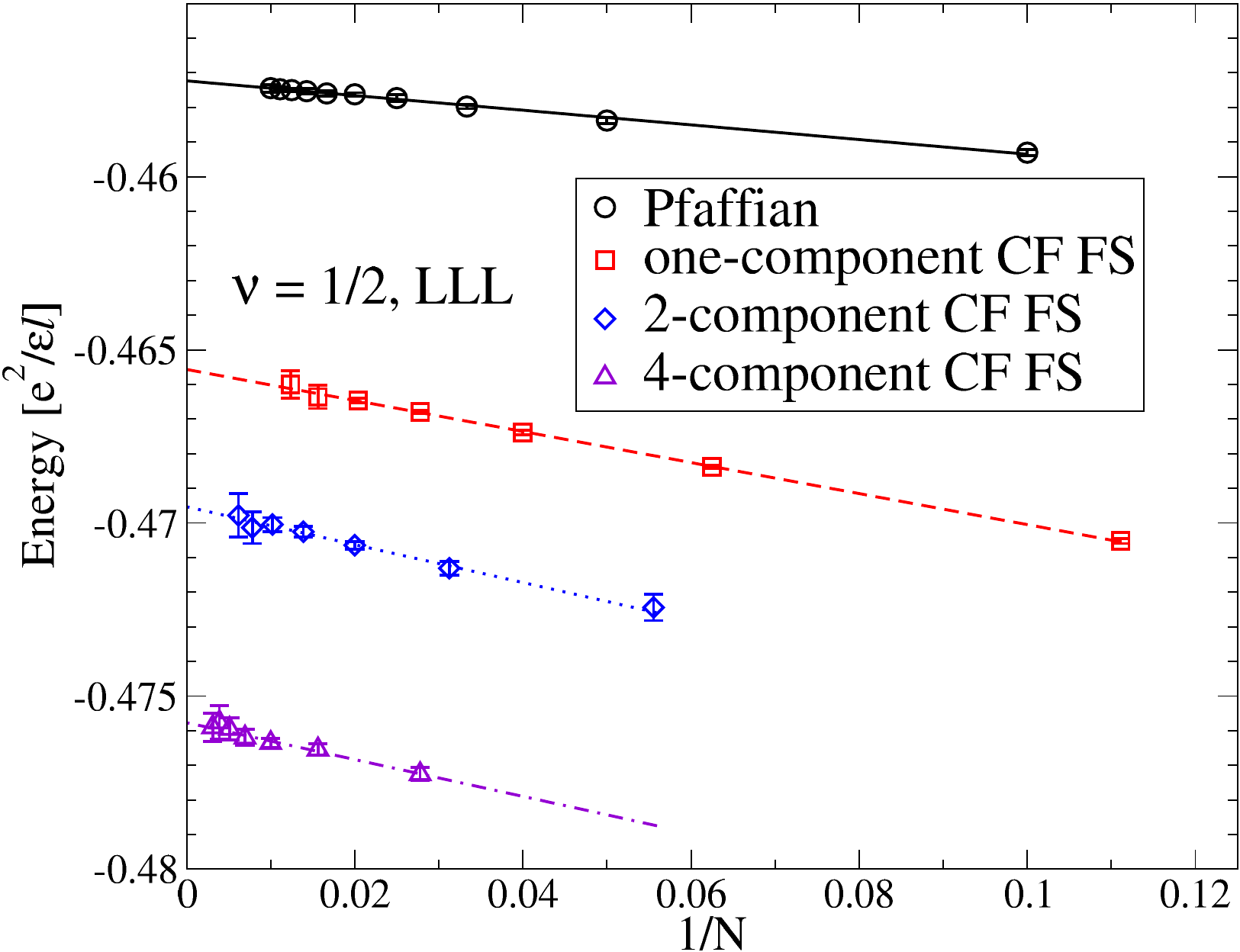}
\includegraphics[width=0.64\columnwidth,keepaspectratio]{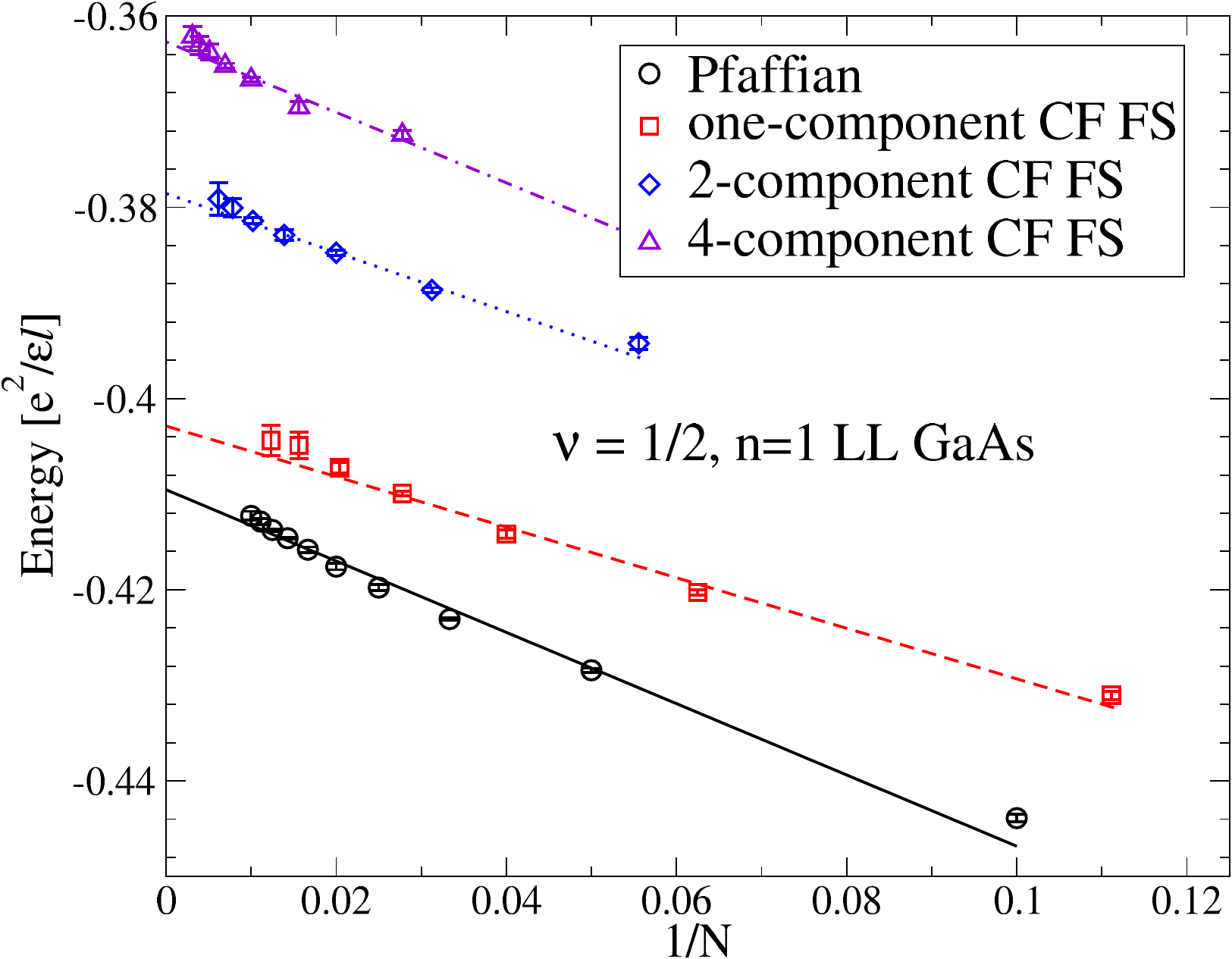}
\includegraphics[width=0.64\columnwidth,keepaspectratio]{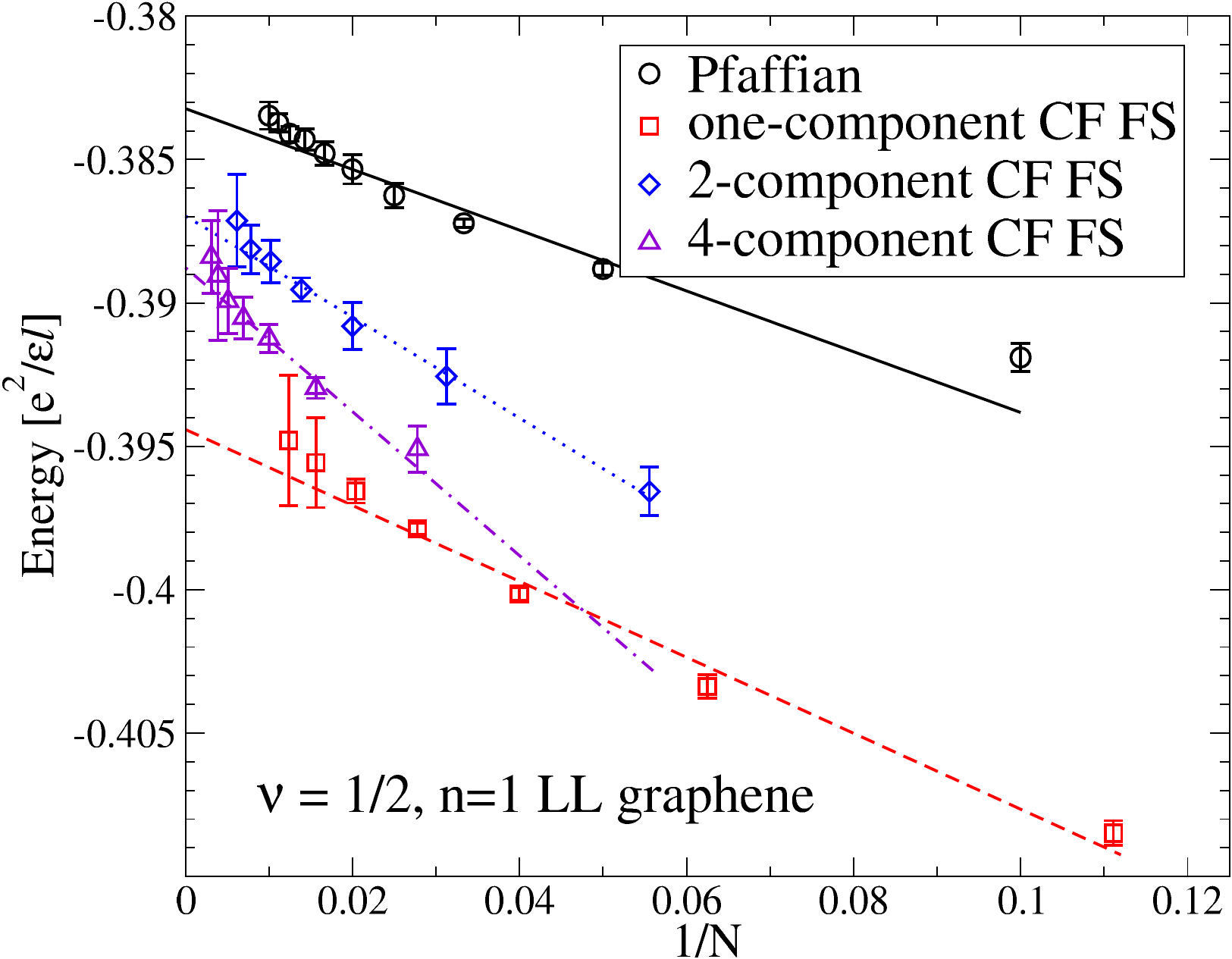}
\end{center}
\caption{\label{nu_1_2}
Extrapolation of the ground state energy to the thermodynamic limit of the states at half filling (a) in the lowest Landau level, (b) in the second ($n=1$) Landau level of GaAs, (c) in the $n=1$ Landau level of graphene. The wave function in Eq.~\ref{Jain_wf} has been used to obtain the energies. The Zeeman energy has been omitted, and the density correction has been applied. The solid lines correspond to a linear fit, and the dashed lines to a quadratic one.
In (b) the energies in the thermodynamic limit may differ from those reported previously in the literature because of different methods for background subtraction, but the energy differences are captured properly. }
\end{figure*}

Fig.\ref{nu_1_2} shows an extrapolation of the energies of the CFFS states (with different spin / pseudospin polarizations) and the MR Pfaffian state. The Pfaffian wave function is clearly ruled out. Furthermore, the fully spin polarized CFFS has the lowest energy. Even though the CF wave function for the spin singlet CFFS is not as accurate as that for the fully polarized CFFS, the energy difference between them is sufficiently large that we are confident in concluding that the ground state at $\nu=1/2$ in the $n=0$ GLL is the fully spin polarized CFFS even at zero Zeeman energy\cite{Wojs11a,Wojs11b}. For contrast, we also show results for two different interactions. One corresponds to the $n=0$ LL, where also the Pfaffian has very high energy, but the ordering of the CFFS states is reversed (with the least polarized state having the lowest energy). The second interaction is that of $n=1$ LL of GaAs, where the Pfaffian wave function is seen to produce the lowest energy among the trial states studied, consistent with a previous study\cite{Park98b}. 

In our studies of $\nu=1/2$ we have neglected the effect of LL mixing. Ref.~\onlinecite{Peterson14} included the effects of LL mixing to the lowest order and found that it favors the anti-Pfaffian state \cite{Levin07,Lee07} (particle-hole conjugate of the Pfaffian state) over the Pfaffian state in the $n=1$ GLL for a certain range of the LL mixing. The question of whether LL mixing can drive the paired state below the fully spin polarized CFFS has not been explored.

Theoretical \cite{Apalkov10,Apalkov11} and experimental \cite{Ki14} evidence exists for the formation of the Pfaffian state in {\em bilayer} graphene. We have focused only on monolayer graphene here. 

\section{Conclusions}
\label{sec:conclusions}

We have used methods of exact diagonalization and CF theory to obtained the ground state energies of differently spin polarized FQHE ground states along the sequence $\nu=s/(2s\pm 1)$ in $n=1$ GLL, as well as their neutral excitations. Our principal conclusion is that these states are fully polarized even at vanishing Zeeman energy. This is in stark contrast to the $n=0$ LL, where, at very small Zeeman energies, the CF state with the smallest polarization is the ground state. We further establish that these states are well described by the CF theory, but the model of weakly interacting composite fermions breaks down in $n=1$ GLL. In the absence of their interaction, composite fermions would occupy the lowest $\Lambda$Ls. When multiple $\Lambda$Ls are filled, the ground state in the absence of Zeeman splitting will be the one with the lowest spin polarization. Hence, the calculated full spin polarization of the actual ground state of electrons with Coulomb interactions implies sufficiently strong interaction among the composite fermions to overcome the single-CF $\Lambda$L splitting (i.e., the effective CF cyclotron energy). Furthermore, in the case of filled $\Lambda$Ls, the relevant interaction among the composite fermions is their exchange. Therefore our calculations indicate that the exchange interaction between composite fermions is sufficiently strong to drive the system ferromagnetic even in the absence of Zeeman coupling. We have also estimated the corrections arising from LL mixing. We find that our conclusions regarding the spin polarization of the FQHE ground state remains unchanged even after incorporating the effect of LL mixing.

We have considered two neutral excitations of the fully polarized ground state at $\nu=s/(2s\pm 1)$: namely the spin-conserving exciton and the spin-flip exciton. We find that the spin-conserving exciton has a dispersion similar to that in the $n=0$ LL. On the other hand, the spin-flip exciton in the $n=1$ GLL does not show any spin-roton with a negative Coulomb energy unlike its counterpart in the $n=0$ LL. This further corroborates the tendency of the ground state to be fully spin polarized at $\nu=s/(2s\pm 1)$. 

Our study further deepens the puzzle surrounding the experimental results of Amet et. al \cite{Amet15} discussed in the Introduction section. They found that the experimental data are sensitive to the Zeeman energy even at very large Zeeman energies, which led them to conclude that the spin was playing a role even at very high magnetic fields. Our study, on the other hand, indicates that the system is fully polarized, at least at the special fractions, even at zero Zeeman energy. An explanation of the Amet {\em et al.} result remains elusive.

Finally we looked at the ground state at half filling in the $n=1$ GLL. Here, we find that CFFS remains stable, and CF pairing is not favored.  Furthermore, the CFFS remains fully spin polarized even at zero Zeeman energy. These facts should be contrasted with the $n=1$ LL of non-relativistic electrons of conventional semiconductors where the CFFS is unstable to a Pffafian type pairing, and the $n=0$ LL of conventional semiconductors where the CFFS is unpolarized in the absence of Zeeman coupling.

{\bf Acknowledgment}
We thank the financial support of U.S. Department of Energy, Office of Science, Basic Energy Sciences, under Award No. DE-SC0005042 (ACB, JKJ),  Hungarian Scientific Research Funds No. K105149 and the Hungarian Academy of Sciences (C.T.), and the National Science Centre, Poland, under grant 2014/14/A/ST3/00654 and the EU Marie Curie Grant PCIG09-GA-2011-294186 (A.W.). We thank Research Computing and Cyberinfrastructure at Pennsylvania State University, which is supported in part through instrumentation funded by the National Science Foundation through grant OCI–0821527, the HPC facility at the Budapest University of Technology and Economics and Wroc{\l}aw Centre for Networking and Supercomputing and Academic Computer Centre CYFRONET, both parts of PL-Grid Infrastructure.

\bibliography{biblio_fqhe}
\bibliographystyle{apsrev}
\end{document}